\documentclass[12pt,onecolumn,floatfix,altaffilletter,superscriptaddress,tightenlines,showpacs,showkeys,preprintnumbers,nofootinbib]{revtex4-1}
\pdfoutput=1

\usepackage[colorlinks=true,citecolor=blue,linkcolor=blue,breaklinks=true]{hyperref}
\usepackage{amsmath,amssymb}
\usepackage{epsfig} 
\usepackage{graphicx}        
\usepackage{url}
\usepackage{color}
\usepackage{multirow}
\usepackage{placeins}
\usepackage[dvipsnames]{xcolor}
\usepackage{braket}
\hypersetup{
  colorlinks=true,
  linkcolor=red,
  filecolor=magenta,   
  urlcolor=blue,
  citecolor=blue,
} 
\providecommand{\xlink}[1]
 {\href{http://arxiv.org/abs/#1}{arXiv:#1}}
 
\allowdisplaybreaks

\bibpunct{[}{]}{,}{n}{}{,}

\def\beq{\begin{equation}}
\def\eeq{\end{equation}}
\def\bea{\begin{eqnarray}}
\def\eea{\end{eqnarray}}
\makeatletter
\@addtoreset{equation}{section}

\begin{document}

\title{Flavoured leptogenesis and ${\rm CP}^{\mu\tau}$ symmetry}

\author{Rome Samanta}
\email{R.Samanta@soton.ac.uk}
\affiliation{Physics and Astronomy, University of Southampton, Southampton, SO17 1BJ, U.K.}
\author{Manibrata Sen}
\email{manibrata@berkeley.edu }
\affiliation{Department of Physics, University of California Berkeley, Berkeley, California 94720, USA.}
\affiliation{Department of Physics and Astronomy, Northwestern University, Evanston, IL 60208, USA.}

\preprint{NUHEP-TH/19-08}

\begin{abstract} 
We present a systematic study of leptogenesis in neutrino mass models with $\mu\tau$-flavoured CP symmetry. In addition to the strong hierarchical $N_1$-dominated scenario ($N_1$DS) in the `two flavour regime' of leptogenesis, we show that one may choose the right-handed (RH) neutrino mass hierarchy as mild as $M_2\simeq 4.7 M_1$ for a perfectly valid hierarchical $N_1$DS. This reduces the lower bound on the allowed values of $M_1$, compared to what is stated in the literature. The consideration of flavour effects due to the heavy neutrinos also translate into an upper bound on $M_1$. It is only below this bound that the observed baryon-to-photon ratio can be realized for a standard ${ N_1}$ domination, else a substantial part of the parameter space is also compatible with $N_2$DS. We deduce conditions under which the baryon asymmetry produced by the second RH neutrino plays an important role. Finally, we discuss another scenario where lepton asymmetry generated by $N_2$ in the two flavour regime faces washout by $N_1$ in the three flavour regime. Considering a hierarchical light neutrino mass spectrum, which is now favoured by cosmological observations, we show that at the end of $N_1$-leptogenesis, the asymmetry generated by $N_2$ survives only in the electron flavour and about $33\%$ of the parameter space is consistent with a pure $N_2$-leptogenesis.

\end{abstract}

\maketitle

\section{Introduction} \label{s1}
%%%%%%%%%%%%%%%%%%%%%%%%%
%%%%%%%%%%%%%%%%%%%%%%%%%
Neutrino masses and mixings continue to intrigue. 
Precise measurement of the six mixing parameters -- the three mixing angles: solar $(\theta_{12})$, atmospheric $(\theta_{23})$ and reactor $(\theta_{13})$, the two mass-squared differences: solar $(\Delta m^2_{12})$, and atmospheric $(\Delta m^2_{23})$, and the CP phase $\delta$ -- is essential for a clear understanding of neutrino physics. 
While we have almost zeroed in on the values of the mixing angles and mass-squared differences from solar, atmospheric and terrestrial experiments \cite{pdg}, we are still pretty much in the dark when it comes to the CP phase. To this end, significant improvements have been made to the experimental determination of $\delta$ in experiments such as T2K \cite{t2k,t2k1,t2k2} and NO$\nu$A \cite{nova1,nova2}. There exists a mild preference for the normal mass ordering (NMO) of the neutrinos, while latest global fit of neutrino oscillation data \cite{globalfit} seem to favour the second octant of $\theta_{23}$ (the best-fit value $\sin^2\theta_{23}=0.58$), and a maximal value of the CP phase $\delta=3\pi/2$ (driven by T2K neutrino and anti-neutrino appearance data) for both the mass orderings. However, precise statements on the mass ordering, octant of $\theta_{23}$ and the value of $\delta$ are yet to be made with a high degree of confidence level. 

This is an exciting time in low energy neutrino phenomenology. Models which have concrete predictions for the yet undetermined parameters such as $\theta_{23}$ and $\delta$ can be tested in the light of recent experimental data. From a theory standpoint, flavour symmetries \cite{Altarelli:2010gt,Ishimori:2010au,King:2017guk,petcov} have always been invoked in neutrino mass models to estimate neutrino mixing parameters. 
A popular example is the $\mu\tau$ symmetry  \cite{m1,m2,m3,m4,Fukuyama:1997ky,Fukuyama:2017qxb, Mohapatra:2005yu,Xing:2006xa}, which was ruled out by the discovery of a non-zero $\theta_{13}$. However, after the hint of maximal CP violation by T2K \cite{t2k}, another variant of the $\mu\tau$ symmetry, the $\mu\tau$ flavoured CP symmetry (${\rm CP}^{\mu\tau}$) or $\mu\tau$ reflection symmetry \cite{cp1,cp2,cp3} has been a topic of interest in the recent years \cite{cp5a,cp6,cp7,cp8,cp9,cp10,cp11,cp12,cp13,cp14,cp15p,cp15,cp16,cp17,cp18,
cp19,cp20,cp21,cp22,cp23}. 

The ${\rm CP^{\mu\tau}}$ symmetry, which is a CP transformation \cite{cp4,cp5,Chen:2015nha} on the left-handed (LH) neutrino fields with $\mu\tau$ interchange symmetry as the CP generator in the low energy effective neutrino Lagrangian, predicts a co-bimaximal mixing \cite{Ma:2015fpa}: $\theta_{23}=\pi/4$ and $\delta=\pi/2,\,3\pi/2$, along with arbitrary non-zero values of $\theta_{13}$. To make ${\rm CP^{\mu\tau}}$ more predictive, a sizable body of work exists to combine flavour symmetries with CP symmetries, despite this being a non-trivial task \cite{cp7,cp8}. Several aspects of ${\rm CP}^{\mu\tau}$ and its alternative versions have also been explored \cite{mtv1,mtv2,mtv3,mtv4,mtv5,mtv6,mtv61,mtv62,mtv7,mtv8}.

From the point of view of cosmology, the ${\rm CP}^{\mu\tau}$ model has generated considerable interest in the possibility of baryogenesis via leptogenesis \cite{lep1,lep2,lep3,lep4,lep5,lep6}. In an extended Standard Model (SM), augmented with right-handed (RH) neutrinos, tiny masses for the active neutrinos can be generated through the Type-I seesaw mechanism \cite{sw1,sw2,sw3}. In such models, CP-violating and out of equilibrium decays of the heavy RH neutrinos can generate a lepton asymmetry (leptogenesis) which can be converted into a baryon asymmetry (baryogenesis) by sphalerons \cite{Kuzmin:1985mm,lep1,lep4}. These sphaleronic transitions conserve ${\rm B-L}$, and violate ${\rm B+L}$, where ${\rm B}$ and ${\rm L}$ are the baryon and lepton number respectively. Given a neutrino mass model, successful baryogenesis requires \cite{planck}
\beq{
\eta_B^{th}\equiv\eta_B^{CMB}=\left(6.3\pm 0.3\right)\times10^{-10}\,,\label{obseta}
}\eeq
where $\eta_B^{th}$ and $\eta_B^{CMB}$ are the theoretical and observed values of baryon to photon ratio at the recombination. Assuming a $N_1$ dominated scenario (${ N_1{\rm DS}}$), where only the decays and interactions of $N_1$ matter, it has been pointed out that ${\rm CP}^{\mu\tau}$ \cite{cp3,cplepto1,hagelepto} as well as the CP symmetries similar to ${\rm CP}^{\mu\tau}$, e.g., CP-anti $\mu\tau$ (${\rm CP}^{\mu\tau A}$) \cite{cp16}, and complex scaling \cite{mtv6,mtv7} are capable of reproducing the observed value of $\eta_B$. This, however, requires the lightest RH neutrino mass to lie within the range $10^9~{\rm GeV}<M_1<10^{12}~{\rm GeV}$ -- so called the two flavour regime (2FR) \cite{fl1,fl2,fl3,fl4} of leptogenesis. For ${\rm CP}^{\mu\tau}$ as well as ${\rm CP}^{\mu\tau A}$, it has also been argued that the regimes $M_1>10^{12}~{\rm GeV}$ -- one flavour regime (1FR) and $M_1<10^{9}~{\rm GeV}$ -- three flavour regime (3FR) -- are not favoured for successful leptogenesis due to the typical structure of the symmetry (we shall discuss it in detail in Sec.\ref{s5}). In the ${ N_1{\rm DS}}$, leptogenesis has been studied with a strong hierarchical scenario \cite{cp3,cplepto1,cp16}, e.g. , $M_2/M_1=10^3$; i.e., assuming other heavy neutrinos are not produced at all, or if produced, the lepton asymmetry due to $N_2$ faces a significant washout by the $N_1$-interactions and thus is negligible, whereas that produced by $N_1$ does not encounter a $N_2$-washout. In addition, a lower bound on $M_1$ has been derived \cite{cp3,cp16} using the neutrino oscillation data and the observed range of $\eta_B$.\\

In this paper we investigate viability of those results in detail. 
After a systematic analysis, we argue the following: 

i) In the ${\rm CP}^{\mu\tau}$ framework, even in the two RH neutrino seesaw model  \cite{King:1999mb,King:2002nf} which is tightly constrained by the neutrino oscillation data, one can choose the heavy RH neutrino mass hierarchy as low as $M_2/M_1\simeq 4.7$ for a perfectly valid hierarchical ${\rm N_1DS}$ leptogenesis scenario. This in turn leads to a decrease in the lower bound on $M_1$, approximately by an order of magnitude. 

ii) Allowing both the RH neutrinos to contribute to the final asymmetry and taking into account the heavy neutrino flavour effects, we show that in the two flavour regime, there is a particular RH neutrino mass window $M^{\rm max}>M_1>M^{\rm min}$ for which the hierarchical ${\rm N_1DS}$ is valid. Beyond $M^{\rm max}$, domination of $N_2$ could also become significant in addition to $N_1$. 
%We comment on the goodness of each $N_i$ domination by calculating $\chi^2$ values of the corresponding parameter space with respect to the best fit values of neutrino oscillation parameters.

iii) Finally, we demonstrate that if the lepton asymmetry is produced by $N_2$ in the two flavour regime and faces washout by $N_1$ in the three flavour regime, then the final asymmetry mainly survives in the electron flavour. This is because the $N_1$-decay parameters for the other two flavours ($K_{1\mu}$ and $K_{1\tau}$) are strong enough to erase any pre-existing asymmetry in the respective flavours. We quantify the probability of $N_2$ leptogenesis to be around $33\%$. This is done by computing the probability of the electron flavour washout parameter $K_{1e}$ to be less than unity, since typically for these values of $K_{1e}$, the asymmetry generated by $N_2$ does not get washed out by $N_1$ \cite{DiBari:2008mp,Blanchet:2011xq,DiBari:2005st,DiBari:2018fvo}\footnote{Following \cite{lep6} we address $K_{1\alpha}$ as decay parameters throughout.}.
 
 The rest of the paper is organized as follows: In Sec.\ref{s2} we briefly discuss the ${\rm CP}^{\mu\tau}$ and its other variants. For simplicity we only focus on the two RH neutrino model, commonly known as the minimal seesaw. Sec.\ref{s3} contains a discussion about the validity of ${\rm N_1 DS}$ in one flavour case which can trivially be generalized into multi flavoured leptogenesis scenario. In Sec.\ref{s4}, we emphasize on the importance of heavy neutrino flavour effects which open up the possibility for $N_2$ leptogenesis. Sec.\ref{s5} contains a thorough discussion of leptogenesis in the model under consideration. We conclude our work in Sec.\ref{s6} emphasizing the main results of this work. 
 %%%%%%%%%%%%%%%%%%%%%%%%
 %%%%%%%%%%%%%%%%%%%%%%%%
 \section{$\rm CP^{\mu\tau}$ symmetry and its variants in seesaw model} \label{s2}
%%%%%%%%%%%%%%%%%%%%%%%% 
 %%%%%%%%%%%%%%%%%%%%%%%%
Before we proceed, we discuss some aspects of the ${\rm CP^{\mu\tau}}$ symmetry in neutrino mass models. Note that we work in a basis where the charged lepton mass matrix $m_\ell$ and the RH neutrino mass matrix $M_R$ are diagonal \cite{cp3,cplepto1}. Thus, the neutrino mixing matrix $U$ can be written as 
\bea
U=P_\phi U_{PMNS}\equiv 
P_\phi \begin{pmatrix}
c_{1 2}c_{1 3} & e^{i\frac{\alpha}{2}} s_{1 2}c_{1 3} & s_{1 3}e^{-i(\delta - \frac{\beta}{2})}\\
-s_{1 2}c_{2 3}-c_{1 2}s_{2 3}s_{1 3} e^{i\delta }& e^{i\frac{\alpha}{2}} (c_{1 2}c_{2 3}-s_{1 2}s_{1 3} s_{2 3} e^{i\delta}) & c_{1 3}s_{2 3}e^{i\frac{\beta}{2}} \\
s_{1 2}s_{2 3}-c_{1 2}s_{1 3}c_{2 3}e^{i\delta} & e^{i\frac{\alpha}{2}} (-c_{1 2}s_{2 3}-s_{1 2}s_{1 3}c_{2 3}e^{i\delta}) & c_{1 3}c_{2 3}e^{i\frac{\beta}{2}}
\end{pmatrix}\,,\nonumber\\
\label{eu}
\eea
where $P_\phi={\rm diag}~(e^{i\phi_1},~e^{i\phi_2}~e^{i\phi_3})$ is an unphysical diagonal phase matrix and $c_{ij}\equiv\cos\theta_{ij}$, $s_{ij}\equiv\sin\theta_{ij}$ with the mixing angles $\theta_{ij}=[0,\pi/2]$. CP violation enters in Eq.\,\ref{eu} through the Dirac phase $\delta$ and the Majorana phases $\alpha$ and $\beta$. For simplicity, we focus on the two RH neutrino model \cite{King:1999mb,King:2002nf}, commonly known as minimal seesaw model \cite{min1,min2,min3,Bambhaniya:2016rbb}. Thus with $m_D$ as the Dirac mass matrix, the neutrino part of the Lagrangian can be written as 
\bea
-\mathcal{L}_{mass}^{\nu,N}= \bar{N}_{Ri} (m_D)_{i\alpha}\nu_{L\alpha}
+\frac{1}{2}\bar{N}_{Ri}(M_R)_{ij} \delta _{ij}N_{Rj}^C 
+ {\rm h.c.}\,, \label{seesawlag}
\eea
with $l_{L\alpha}=\begin{pmatrix}\nu_{L\alpha} & e_{L\alpha}\end{pmatrix}^T$ as the SM lepton doublet of flavor $\alpha$ and $M_R={\rm diag}\hspace{.5mm} (M_1,M_2)$, $M_{1,2}>0$. The effective light neutrino mass matrix is given by the standard seesaw relation
\bea
M_\nu = -m_D^TM_R^{-1}m_D\,. \label{seesaweq}
\eea
Now a CP transformation \cite{cp4,cp5} on the LH neutrino field, $\nu_{Ll}\rightarrow iG_{lm}\gamma^0\nu_{Lm}^C\,$, leads to the following invariance of the effective light neutrino mass matrix $M_\nu$:
\bea
G^TM_\nu G=M_\nu^*\,,\label{cpmutau}
\eea
where $G$ is the generator matrix.
If $G$ follows a $\mu\tau$-interchange symmetry \cite{cp1,cp2}, i.e.,
\bea
G=\begin{pmatrix}
1&0&0\\0&0&1\\0&1&0
\end{pmatrix},\label{gmutau}
\eea 
 then the symmetry transformation in Eq.\,\ref{cpmutau} is known as a $\mu\tau$ flavoured CP transformation or ${\rm CP}^{\mu\tau}$ \cite{cp3}. A simple alteration of ${\rm CP}^{\mu\tau}$ has recently been studied by one of the authors, by adding a minus sign to the right hand side of Eq.\,\ref{cpmutau} \footnote{Note that the high energy symmetry could be very different than ${\rm CP}^{\mu\tau}$, as pointed out in \cite{cp16}. }. This symmetry, named as the CP anti-$\mu\tau$ or ${\rm CP}^{\mu\tau A}$\cite{cp16}, could be recast as a symmetry transformation equation similar to Eq.\,\ref{cpmutau} as
\bea
\mathcal{G}^TM_\nu \mathcal{G}=M_\nu^*\,,\label{cpmutaua}
\eea
with $\mathcal{G}=iG$. Intriguingly, the $\mu\tau$ symmetry ($G$) and the $\mu\tau$ antisymmetry ($\mathcal{G}$) have completely different predictions when they are used as an ordinary field transformation, i.e., $\nu_{Ll}\rightarrow G_{lm}\nu_{Lm}$ or $\nu_{Ll}\rightarrow \mathcal{G}_{lm}\nu_{Lm}$ \cite{Grimus:2005jk}.
 However in their CP-transformed versions, along with the diagonalization condition $U^TM_\nu U=M_d$, where $M_d={\rm diag}(m_1,m_2,m_3)$, both the symmetries (Eq.\,\ref{cpmutau} and Eq.\,\ref{cpmutaua}) lead to the same predictions \cite{cp3,cp16}
\beq
\cos\delta =\sin\alpha=\sin\beta=0\,,\qquad
\theta_{23}=\pi/4\,.\label{cputpred}
\eeq
This is easy to understand. Consider a mass matrix $M_\nu$, which follows Eq.\,\ref{cpmutaua}. This can be written in the form
\bea
M_\nu^{{CP}^{\mu\tau A}}=\begin{pmatrix}
iA&B&-B^*\\B&C&iD\\-B^*&iD&-C^*
\end{pmatrix}\,,\label{Amutau}
\eea
where $A,D$ are real and $B,C$ are complex mass dimensional quantities which are a priori unknown. Now mass matrix in Eq.\,\ref{Amutau} also satisfies the equation
\bea
G^T(i M_\nu^{{CP}^{\mu\tau A}})G=(i M_\nu^{{CP}^{\mu\tau A}})^*\,,
\eea
which is basically a ${\rm CP}^{\mu\tau }$ transformation (Eq.\,\ref{cpmutau}). Thus, if a mass matrix follows ${\rm CP}^{\mu\tau A}$ invariance, `$i$' times the same matrix also obeys ${\rm CP}^{\mu\tau }$ symmetry, and hence both the symmetries lead to similar phenomenological consequences. Henceforth, without lack of generality, we shall consider the CP-antisymmetric parametrization of $m_D$ as well as $M_\nu$ derived in \cite{cp16}. 

For a diagonal $M_R$, Eq.\,\ref{cpmutaua} is satisfied through the symmetry transformation on $m_D$ as\footnote{We shall refer the reader Refs. \cite{cp3,mtv7,cplepto1} to have a look to realize how in the diagonal basis of $m_\ell$ and $M_R$, CP symmetry could be applied in the neutrino mass terms.} 
\bea
m_D\mathcal{G}=-im_D^*\,.\label{mdtr}
\eea
The most general form of $m_D$ that satisfies (\ref{mdtr}) can be parametrized as
\bea
m_D=\begin{pmatrix}
\sqrt{2}a_1e^{i\pi/4}&b_1e^{i\theta_1}&ib_1e^{-i\theta_1}\\
\sqrt{2}a_2e^{i\pi/4}&b_2e^{i\theta_2}&ib_2e^{-i\theta_2}\\
\end{pmatrix}\,, \label{mdseesaw}
\eea
where the parameters $a_{1,2}$, $b_{1,2}$ and $\theta_{1,2}$ are real. Now using Eq.\,\ref{seesaweq}, the effective light neutrino mass matrix $M_\nu$ can be written as
\bea
M_\nu^{{CP}^{\mu\tau A}}=\hspace{16cm}\nonumber\\
\begin{pmatrix}
-2i(x_1^2+x_2^2)&-\sqrt{2}e^{i\pi/4}(x_1y_1e^{i\theta_1}+x_2y_2e^{i\theta_2})&-i\sqrt{2}e^{i\pi/4}(x_1y_1e^{-i\theta_1}+x_2y_2e^{-i\theta_2})\\-\sqrt{2}e^{i\pi/4}(x_1y_1e^{i\theta_1}+x_2y_2e^{i\theta_2})&-(e^{2i\theta_1}y_1^2+e^{2i\theta_2}y_2^2)&-i(y_1^2+y_2^2)\\-i\sqrt{2}e^{i\pi/4}(x_1y_1e^{-i\theta_1}+x_2y_2e^{-i\theta_2})&-i(y_1^2+y_2^2)&e^{-2i\theta_1}y_1^2+e^{-2i\theta_2}y_2^2 
\end{pmatrix}.\nonumber\\\label{mnuseesaw}
\eea

In (\ref{mnuseesaw}), new real parameters $x_{1,2}$ and $y_{1,2}$ are defined by scaling $a_{1,2}$ and $b_{1,2}$ with the square roots of the respective ${\rm RH}$ neutrino masses $M_{1,2}$, i.e.
\bea
\frac{a_{1,2}}{\sqrt{M_{1,2}}}= x_{1,2}\,,~~~\frac{b_{1,2}}{\sqrt{M_{1,2}}}= y_{1,2}\,.\label{primed}
\eea
A few comments on the matrix $M_\nu^{{CP}^{\mu\tau A}}$ are in order. Since ${\rm det}~(M_\nu^{{CP}^{\mu\tau A}})=0$, the lightest neutrino mass (either $m_1$ for a normal mass ordering or $m_3$ for an inverted mass ordering) has to vanish. Furthermore, one of the phases in $M_\nu^{{CP}^{\mu\tau A}}$ (say $\theta_1$) could be rotated with the phase matrix $P_\phi={\rm diag}~(1,e^{i\phi},e^{-i\phi})$ by the choice $\theta_1=-\phi$. Therefore, we are left only with the phase difference $\theta_2-\theta_1$, which can be renamed as $\theta$. Without loss of generality, this is equivalent to the choice $\theta_1=0$ and $\theta_2=\theta$ in $m_D$. For phenomenological analysis, we use this redefined phase $\theta$ for both $M_\nu^{{CP}^{\mu\tau A}}$ as well as $m_D$.\\

%%%%%%%%%%%%%%%
%%%%%%%%%%%%%%%
\section{Validity of ${\rm N_1DS}$ in one flavour thermal leptogenesis} \label{s3}
%%%%%%%%%%%%%%%
%%%%%%%%%%%%%%%
In this section, we start by discussing the standard $N_1$ dominated leptogenesis ($\rm N_1DS$) scenario in the presence of another heavy neutrino $N_2$, assuming both of them are thermally produced \cite{Giudice:1999fb,Khlopov:1984pf,Giudice:2003jh,Croon:2019dfw} so that the reheating temperature $T_{\rm RH}>M_{1,2}$. To begin with, we focus on the one-flavour scenario (i.e., no charged lepton flavour effects). The overall conclusions drawn from one flavour approximation can easily be generalized in the presence of flavour effects, as we discuss later. The set of classical kinetic equations \cite{lep6} relevant for leptogenesis could be written as 
\bea
\frac{dN_{N_i}}{dz}&=&-D_i(N_{N_i}-N_{N_i}^{\rm eq}), \quad{\rm with}~i=1,2 \,,\label{be01}\\
\frac{dN_{B-L}}{dz}&=&-\sum_{i=1}^2\varepsilon_i D_i(N_{N_i}-N_{N_i}^{\rm eq})-\sum_{i=1}^2W_iN_{B-L}\,,\label{be02}
\eea
with $z=M_1/T$. The $N_i$'s and $N_{B-L}$ are the abundances per $N_1$'s in ultra relativistic thermal equilibrium. The equilibrium abundances of $N_i$'s are given by $N_i^{\rm eq}=\frac{1}{2} z_i^2\mathcal{K}_2(z_i)$, where $\mathcal{K}_2(z_i)$ are the modified Bessel functions. The total CP asymmetry is quantified by $\varepsilon_i=\sum_{\alpha}\varepsilon_{i\alpha}$ where
 \bea
 \varepsilon_{i\alpha}&=&\frac{\Gamma_{i\alpha}-\bar{\Gamma}_{i\alpha}}{\Gamma_i+\bar{\Gamma}_i} .\label{epsi}
 \eea
 The flavoured CP asymmetry parameter $\varepsilon_{i\alpha}$ can be estimated as
 \bea
\varepsilon_{i\alpha}
&=&\frac{1}{4\pi v^2 h_{ii}}\sum_{j\ne i} {\rm Im}\{h_{ij}
({m_D})_{i\alpha} (m_D^*)_{j\alpha }\}%\nonumber\\
\left[f(x_{ij})+\frac{\sqrt{x_{ij}}(1-x_{ij})}
{(1-x_{ij})^2+{{h}_{jj}^2}{(16 \pi^2 v^4)}^{-1}}\right]\nonumber\\
&+&\frac{1}{4\pi v^2 {h}_{ii}}\sum_{j\ne i}\frac{(1-x_{ij})
{\rm Im}\{{h}_{ji}({m_D})_{i\alpha} (m_D^*)_{j\alpha}\}}
{(1-x_{ij})^2+{{h}_{jj}^2}{(16 \pi^2 v^4)}^{-1}},\label{ncp}
%\label{epsi_intro_h}
\eea 
where $h_{ij}\equiv( m_D m_D^\dag)_{ij}$, $\langle\phi^0\rangle=v/\sqrt{2}$, $x_{ij}=M_j^2/M_i^2$ and $f(x_{ij})$ has the standard expression \cite{cp16}. The decay parameter is given by 
\beq
K_i\equiv\frac{\Gamma_{D,i}(T=0)}{H(T=M_i)}\,,
\eeq
where $H(T=M_i)$ is the Hubble paramter defined at the temperature $T=M_i)$. Using $z_i=z\sqrt{x_{1i}}\,$, the decay terms can be written as
\bea
D_i=\frac{\Gamma_{D,i}}{H z}=K_ix_{1i}z \langle 1/\gamma_i\rangle\,,
\eea
where the total decay rates $\Gamma_{D,i}=\bar{\Gamma}_i+\Gamma_i=\Gamma_{D,i}(T=0) \langle 1/\gamma_i\rangle$ with $ \langle 1/\gamma_i\rangle$'s as the thermally averaged dilution factors given by the ratios of two modified Bessel functions
\bea
 \langle 1/\gamma_i\rangle=\frac{\mathcal{K}_1(z_i)}{\mathcal{K}_2(z_i)}\,.
\eea
The washout factor $W_i$ typically contains three terms: The inverse decay term $W^{\rm ID}_i$, the $\Delta L=1$ scattering term $W^{\Delta L=1}_i$, and the nonresonant part of the $\Delta L=2$ term $W^{\Delta L=2}_i$. For a strong washout scenario \footnote{We show later that a strong wash-out scenario is preferred in the model under consideration.} and hierarchical light neutrino masses, the scattering terms and the $\Delta L=2$ terms can be safely neglected \cite{Buchmuller:2003gz,Buchmuller:2002rq,lep3}. Thus, the relevant washout term $W_i\simeq W_i^{\rm ID}$ can be written as (after properly subtracting the real intermediate state contribution of $\Delta L=2$ process \cite{lep5})
\bea
W_i^{\rm ID}=\frac{1}{4}K_i\sqrt{x_{1i}}\mathcal{K}_1(z_i)z_i^3\,.\label{inv_decay}
\eea
The final $B-L$ asymmetry could be written as
\bea
N_{B-L}^f=N_{B-L}^{\rm in} e^{-\sum_{i}\int dz^\prime W_i(z^\prime)}+N_{B-L}^{\rm lepto}\,,\label{flepto}
\eea
where $N_{B-L}^{\rm in}$ could be a possible pre-existing asymmetry \cite{grav1,grav2} at an initial temperature $T_{\rm in}$. However in this work, we do not consider any possible pre-existing asymmetry which would impose additional constraints\footnote{By ``pre-existing asymmetry" we mean asymmetries that may originate not only from heavier RH neutrinos but also from other external sources. These asymmetries may be large in magnitude. Therefore one needs several conditions on the flavoured decay parameters to washout the pre-existing asymmetry. In literature, somtimes these conditions are referred to as ``strong thermal conditions". See e.g., Ref. \cite{st1,st2}.  } on the model parameter space \cite{Giudice:1999fb,Bertuzzo:2010et}. In fact, as we shall discuss, given the RH neutrino masses in our model, $10^{9}{\rm GeV}<M_1,M_2<10^{12}{\rm GeV}-$2FR, it is not possible to washout a pre-existing asymmetry which is orthogonal to the direction of $N_1$-washout \cite{Engelhard:2006yg,Antusch:2011nz}. Thus, the scenario of a pure leptogenesis from RH neutrino decay  breaks down. Assuming standard thermal history of the universe, the final baryon-to-photon ratio can be written as
\bea
\eta_B=a_{\rm sph}\frac{N_{B-L}^{\rm lepto}}{N_\gamma^{\rm rec}}\simeq 0.96\times 10^{-2}N_{B-L}^{\rm lepto}\,,
\eea
where $N_\gamma^{\rm rec}$ is the normalised photon density at the recombination and the sphaleron conversion coefficient $a_{\rm sph}\sim 1/3$. This theoretically calculated value of $\eta_{B}$ has to be compared with measured value given in Eq.\ref{obseta}.\\

%\bea
%\eta_{B}^{\rm CMB}=(6.3\pm 0.3)\times 10^{-10} \,. 
%\eea
Before discussing validity of the $N_1$DS in presence of $N_{i(i\neq 1)}$, let us introduce another important parameter $\delta_{1i}=(M_i-M_1)/M_1$ which accounts for the mass difference between $M_i$ and $M_1$. This is related to $x_{1i}$ as
\bea
\sqrt{x_{1i}}=1+\delta_{1i}\Rightarrow z_i=z(1+\delta_{1i})\,.
\eea
Armed with all the necessary prerequisites, we solve Eqs.\,\ref{be01}, and \ref{be02} for $N_1$ in the presence of washouts due to both $N_1$ and $N_2$. Note that for the fixed values of $z$ and $K_2$, the strength of the $N_2$-washout ($W^{\rm ID}_2$) depends on $\delta_{21}$ (cf. Eq.\,\ref{inv_decay}). As a result, solutions of Eq.\,\ref{be02} for different values of $\delta_{12}$ 
indicates a minimum, below which the effect of $W^{\rm ID}_2$ starts to become prominent. This helps to reproduce the standard hierarchical $N_1$ dominated scenario.
 
\begin{figure}[!t]
\includegraphics[scale=0.35]{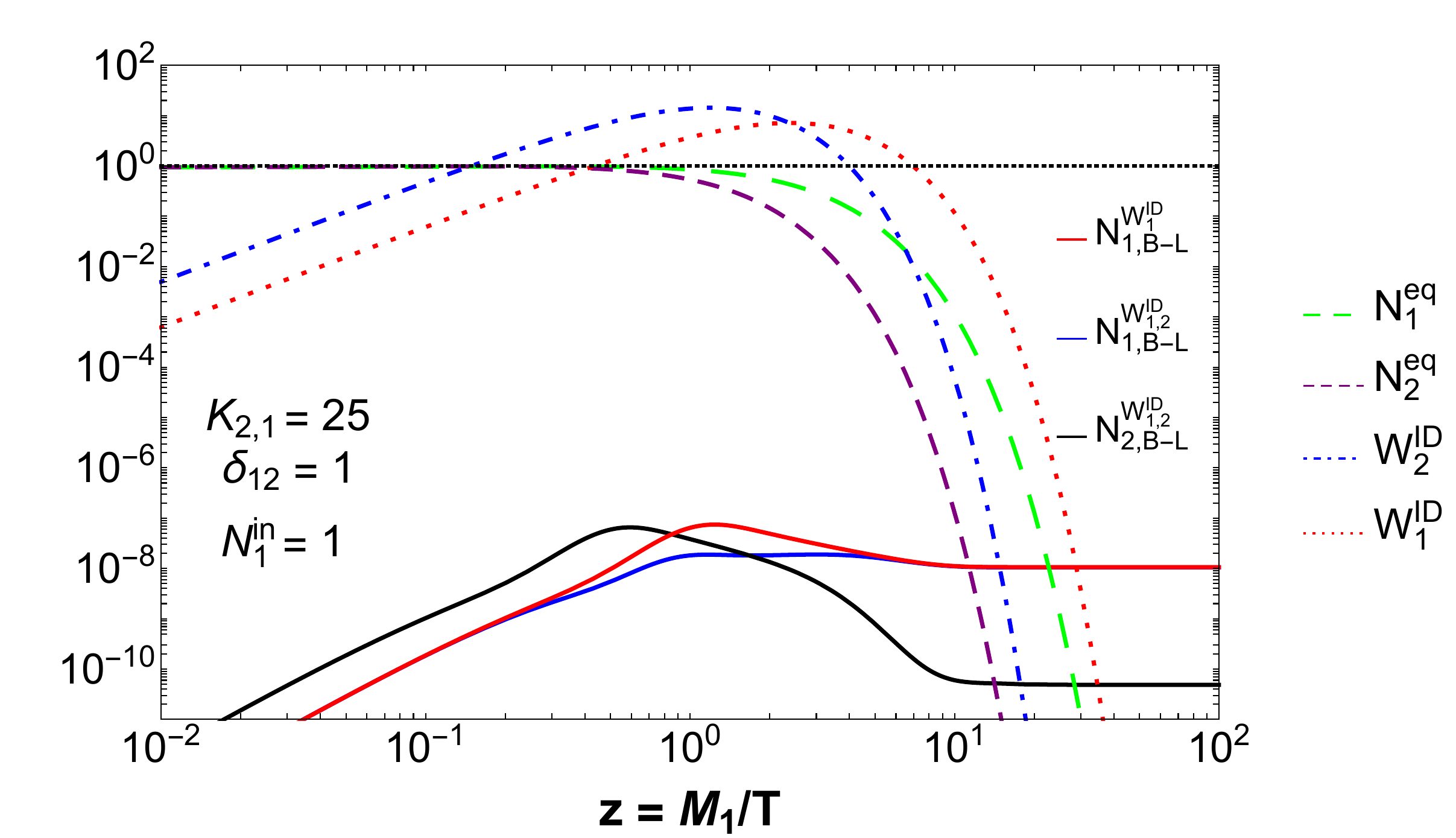}\includegraphics[scale=0.35]{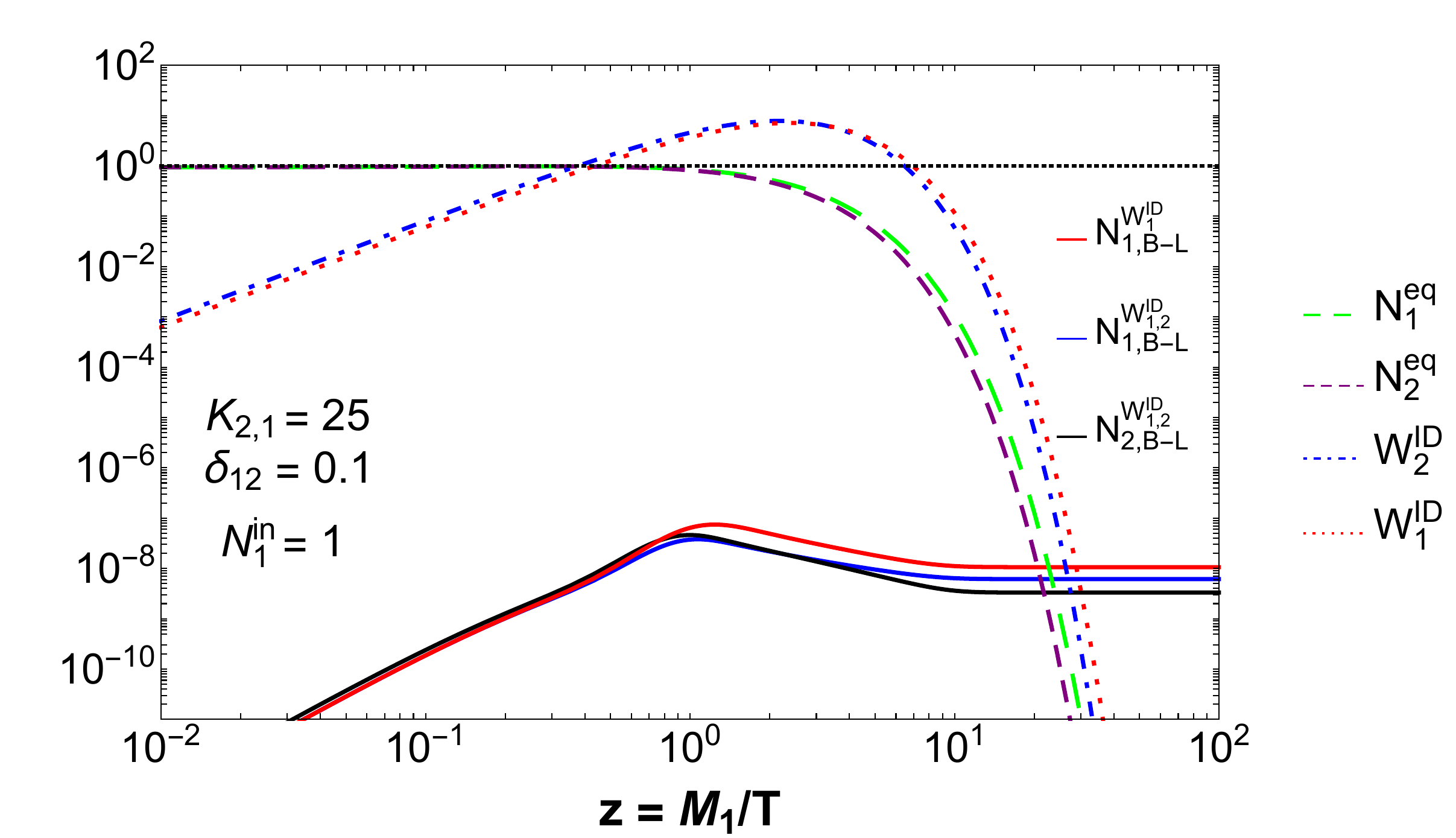}\\
\includegraphics[scale=0.35]{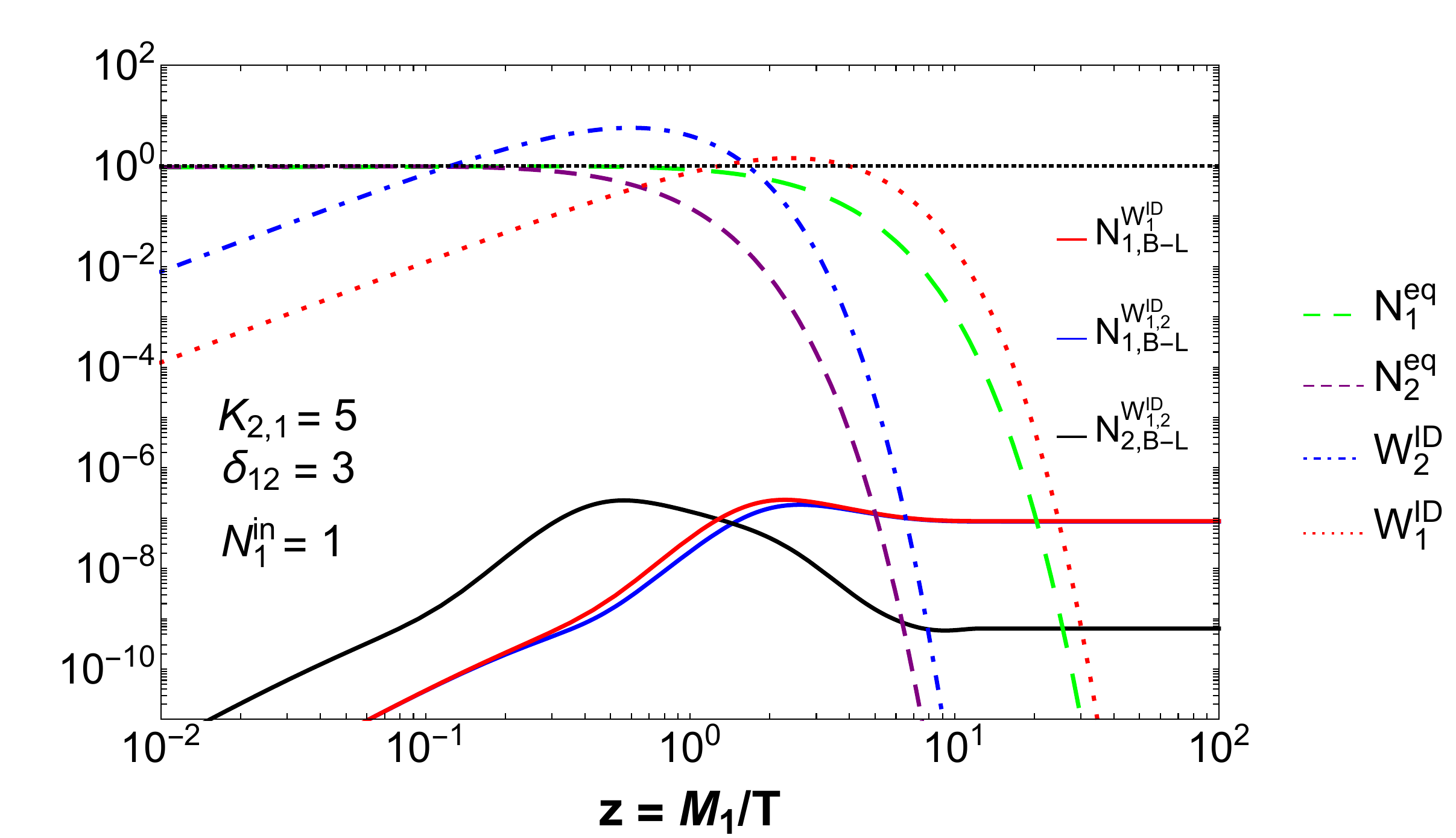}\includegraphics[scale=0.35]{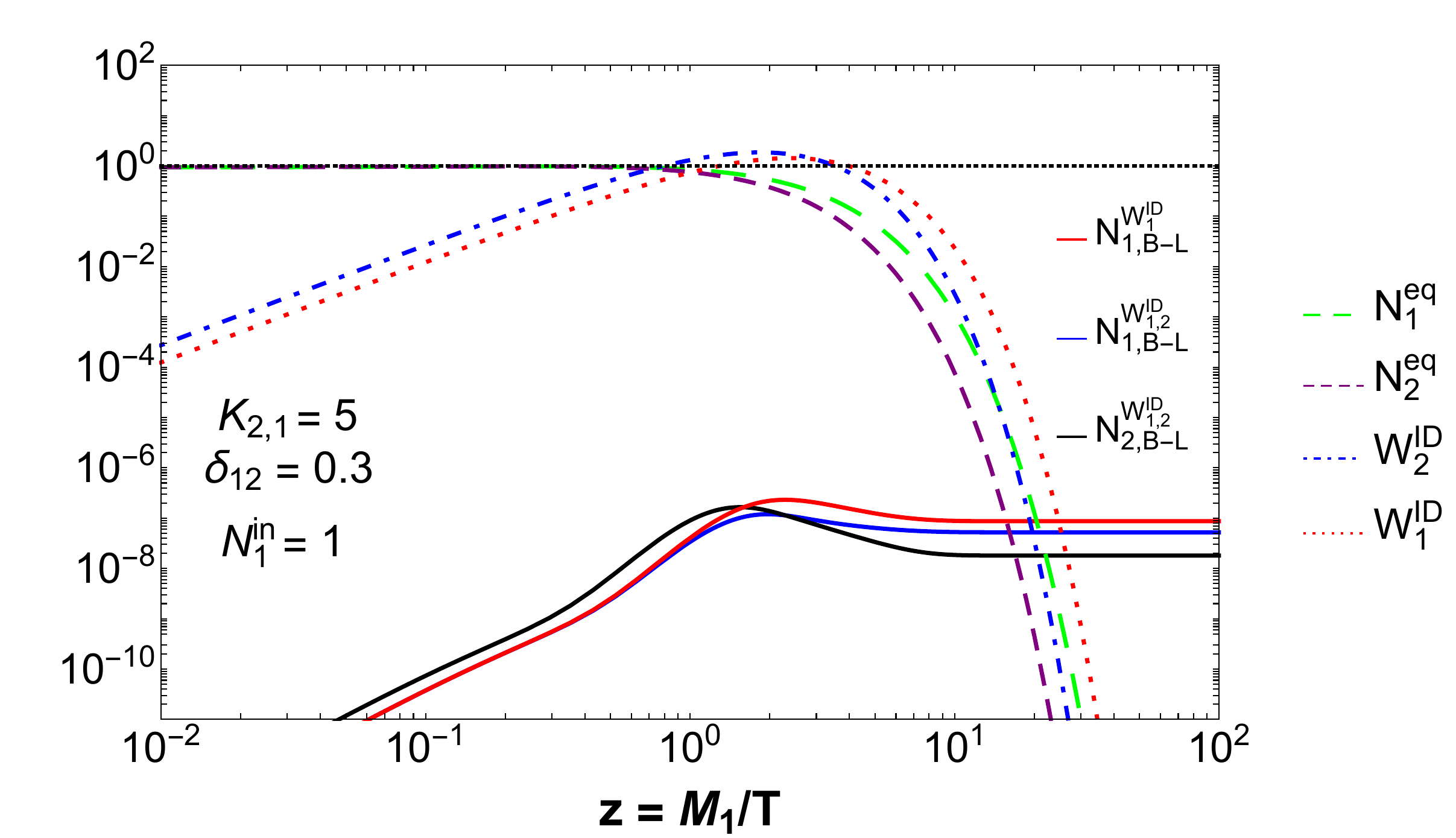}
\caption{Top left: $|N_{B-L}|$ as a function of $z=M_1/T$ for the decay parameters $K_1=K_2=25$ (the other relevant quantities, e.g., $N_{1}^{\rm eq}$, $W_1^{\rm ID}$ etc., are mentioned on the right side of each figure). Solid red line shows $|N_{B-L}|$ for a pure $N_1$ dominated scenario. The solid blue line shows the asymmetry generated by $N_1$, for $\delta_{12}=1$, subjected to both $N_1$ and $N_2$ washout , given by $W_{1,2}^{\rm ID}$ respectively. The solid black line shows $N_{B-L}$, generated by $N_2$ subjected to $W_{1,2}^{\rm ID}$ washout. Top right: For the same value of the decay parameters we generate similar plots for $\delta_{12}=0.1$. Bottom panel shows similar plots as those in the top panel for $\delta_{12}=3$ (left) and $\delta_{12}=0.3$ (right), for $K_1=K_2=5$. 
}\label{f1}
\end{figure}
In Fig.\,\ref{f1}, we show the variation of the produced asymmetry, $|N_{B-L}|$, with $z$. In each figure, $N_{B-L}$ lines in red and blue correspond to the asymmetry produced by $N_1$ subjected to $N_1$, and $N_1+N_2$ washout respectively. The asymmetry showed in black is that produced by $N_2$ subjected to $N_1+N_2$ washout. Figures in the top and bottom panel are for $K_1=K_2=25$ and $K_1=K_2=5$ respectively. 

It is clear from the top-left panel that for $\delta_{12}=1$, even if one takes into account the $N_2$ washout alongwith the $N_1$ washout, the final asymmetry perfectly coincide with standard $N_1$DS. This is simply because the $N_2$ washout goes out of equilibrium before the asymmetry production due to $N_1$ stops. Thus, the final dynamics is governed by the inverse decays of $N_1$ (i.e., $N_1$ washout). On the other hand, the asymmetry produced by $N_2$ is significantly washed out by $N_1$ (showed in black). This is due to the fact that when the strength of the $N_1$ inverse decay reaches its maximum value, the asymmetry production due to $N_2$ is practically switched off. On the top-right panel, we show the same quantities, but for $\delta_{12}=0.1$. Note that in this case, there is a clear distinction between a pure $N_1$ dominated scenario, and that where $N_2$ washout is also taken into account. Here, the $N_2$ washout of the asymmetry production due to $N_1$ cannot be ignored, and hence, the magnitude of $N_{B-L}$ reduces. Furthermore, the $N_1$ inverse decay cannot fully washout the asymmetry produced by $N_2$, since even when the $N_1$ washout is significant, asymmetry production due to $N_2$ does not cease. This causes a significant increase in the magnitude of the asymmetry produced by $N_2$. 
The bottom panel shows the same plots for $K_1=K_2=5$. In this case, however, pure $N_1$DS is realised with slightly increased value of $\delta_{12}=3$, as opposed to $\delta_{12}=1$. For completeness, we also show the plots with $\delta_{12}=0.3$ for which one cannot assume a pure $N_1$DS due to the crucial role played by $N_2$. 

This begs the following question: what is the minimum hierarchy in the RH neutrino masses so that a pure hierarchical $N_1$DS is realized? For example, as discussed, if some model predicts a simple correlation between the decay parameters, say, $K_1=K_2 \in (5-25)$, one can safely assume $\sqrt{x_{12}}=M_2/M_1=(1+\delta_{12})=4$, so that the effect of $N_2$ washout at $N_1$-leptogenesis phase, as well as the asymmetry produced by $N_2$, can be neglected. However, for a realistic scenario, the correlation of the decay parameters may not be this simple; also, a realistic model might contain lots of data points constrained by neutrino oscillation data. Thus in terms of computation, it would be tedious to solve Boltzmann equations for each and every pair of decay parameters. It is useful then to consider explicit and accurate analytic formalism for the computation of these parameters \cite{lep3,lep6,fl1}. To this end, we use the analytic formulae outlined in \cite{lep6}. We first do a consistency check of the results that we discussed after solving the Boltzmann equations with those obtained by the analytic formulae. Then we briefly discuss the overall implementation procedure of the analytic solutions that will be followed in the context of the concerned model.\\

Solutions to Eq.\,\ref{be01} and \ref{be02} can be written as \cite{lep4}
\bea
N_{B-L}^{\rm lepto}=-\sum_{i}^2\varepsilon_{i}\kappa_i\,,
\eea
where $\kappa_i$ is the efficiency of the asymmetry production due to the $i^{\rm th}$ RH neutrino and is given by
\bea
\kappa_i (z)=-\int_{z_{\rm in}\rightarrow 0}^{z_{\rm fin}\rightarrow\infty} \frac{dN_{N_i}}{dz^\prime}e^{-\sum_{i}\int_{z^\prime}^z W_{i}^{\rm ID}(z^{\prime\prime})dz^{\prime\prime}}dz^\prime\,.\label{effi1}
\eea
For a strong washout regime, $ \frac{dN_{N_i}}{dz^\prime}\simeq \frac{dN_{N_i}^{\rm eq}}{dz^\prime}$, since the Yukawa couplings are strong enough to let any species of $N_i$ reach the equilibrium density, even if one starts from vanishing thermal abundance. 
%For practical computation one can choose $z_{\rm in}$ close to `zero' and $z$ to be a large number (theoretically, $\infty$). 
 One has to compare the $\kappa_i (z\rightarrow\infty)$, obtained by solving Eq.\,\ref{effi1} numerically, with the efficiency factor $\kappa_i^{\infty}$, obtained for a pure $N_1$ or $N_2$ dominated scenario, calculated at $z\rightarrow \infty$ and for thermal initial abundances of the RH neutrinos \cite{lep6},\footnote{In any case, for strong washout regime, final asymmetry does not depend upon initial conditions, e.g., see \cite{lep3,lep6}.}
\bea
\kappa_1^\infty &=&\frac{2}{K_1 z_B (K_1)} \left(1-e^{-\frac{K_1 z_B(K_1)}{2}}\right),\\
\kappa_2^\infty &=& \frac{2}{K_2 z_B (K_2)} \left(1-e^{-\frac{K_2z_B(K_2)}{2}}\right)e^{-\int_{0}^\infty W_1^{\rm ID}(z)dz}\,,\nonumber\\ &= & \frac{2}{K_2 z_B (K_2)} \left(1-e^{-\frac{K_2 z_B(K_2)}{2}}\right) e^{-3\pi K_1/8 }\,,
\eea
where 
\bea
z_{B}(K_i)=2+4K_i^{0.13} e^{-\frac{2.5}{K_i}}\,.
\eea
\begin{figure}[!t]
\begin{center}
\includegraphics[scale=0.65]{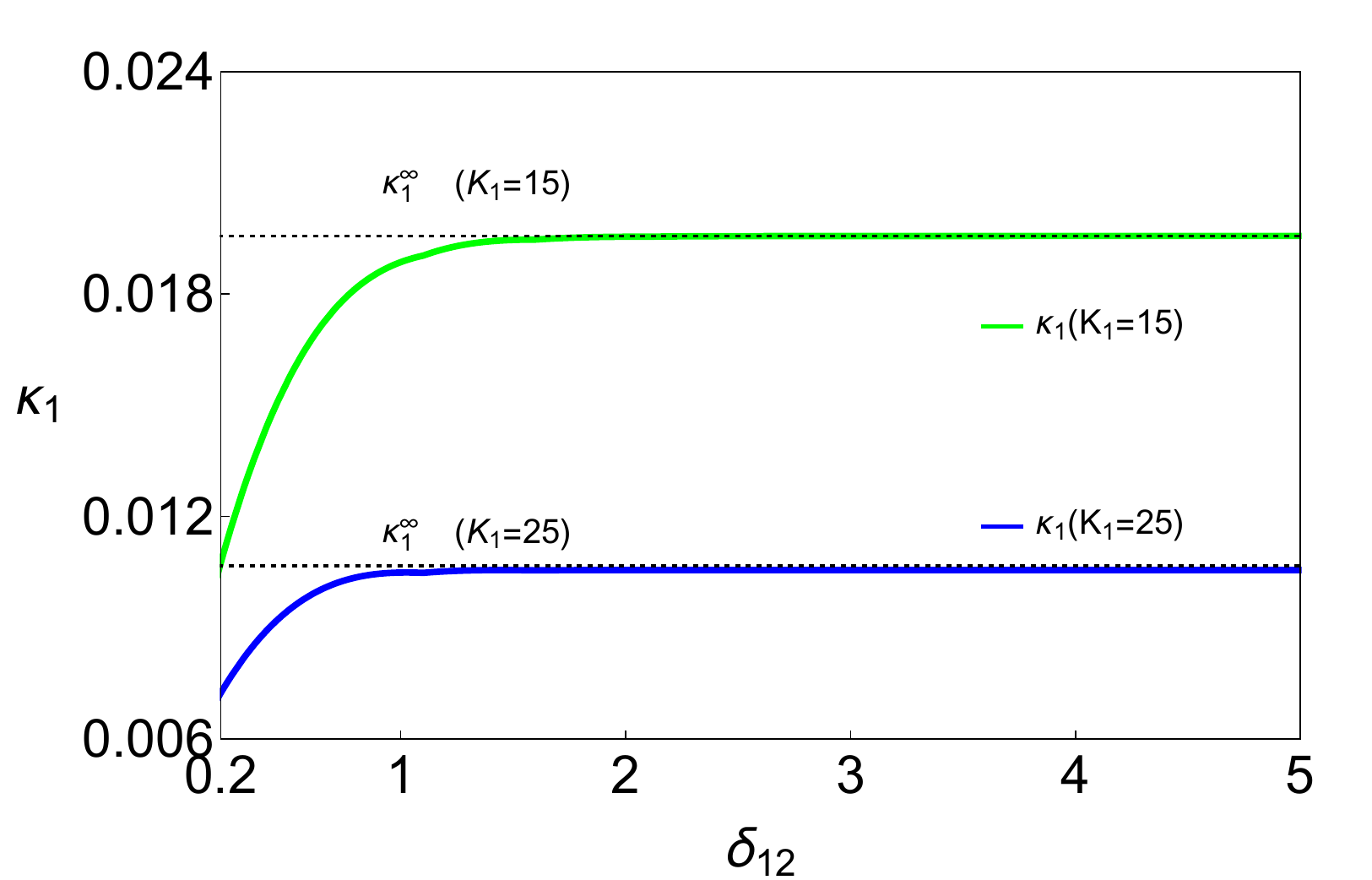}
\caption{ Efficiency factor with $\delta_{12}$ for two different values of $K_1$ with a fixed value of $K_2=25$.
 }\label{f2}
\end{center}
\end{figure}
To arrive at the exponential washout of $\kappa_2^\infty$ by $N_1$, we use
\bea
\int_{0}^\infty z^{\alpha -1} \mathcal{K}_n(z) dz= 2^{\alpha -2 }\Gamma\left( \frac{\alpha -n}{2}\right)\Gamma\left( \frac{\alpha +n}{2}\right)\,.
\eea

 In Fig.\,\ref{f2}, we show the comparison between $\kappa_{1}$ and $\kappa_{1}^{\infty}$ for two different values of $K_{1,2}\in (15,\,25)$. We find that for $K_{1,2}=25$, there is an excellent match between $\kappa_1$ and $\kappa_1^\infty$ for $\delta_{12}\geq1$, which is consistent with the conclusions drawn in Fig.\,\ref{f1} (top-left). However, as expected, when one considers a lower value for $K_1$, say $K_1=15$, it is no longer safe to use $\delta_{12}=1$ for a hierarchical $N_1$DS. 
 %This is because in this case, the strength of the $N_1$ inverse decay starts to decrease and $N_2$ washout affects the asymmetry production more significantly. 
 
\begin{figure}[!t]
\includegraphics[scale=0.48]{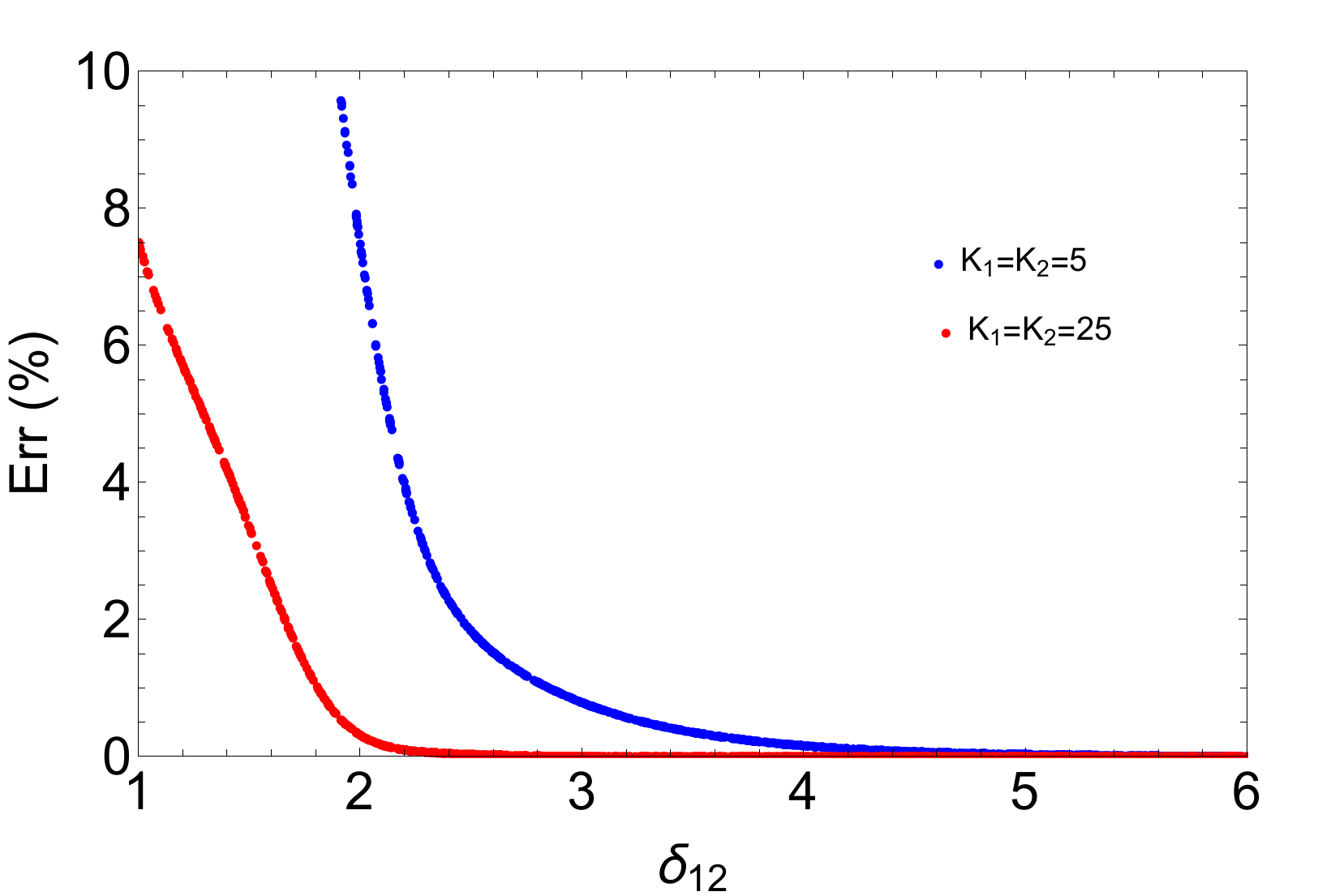} \includegraphics[scale=0.44]{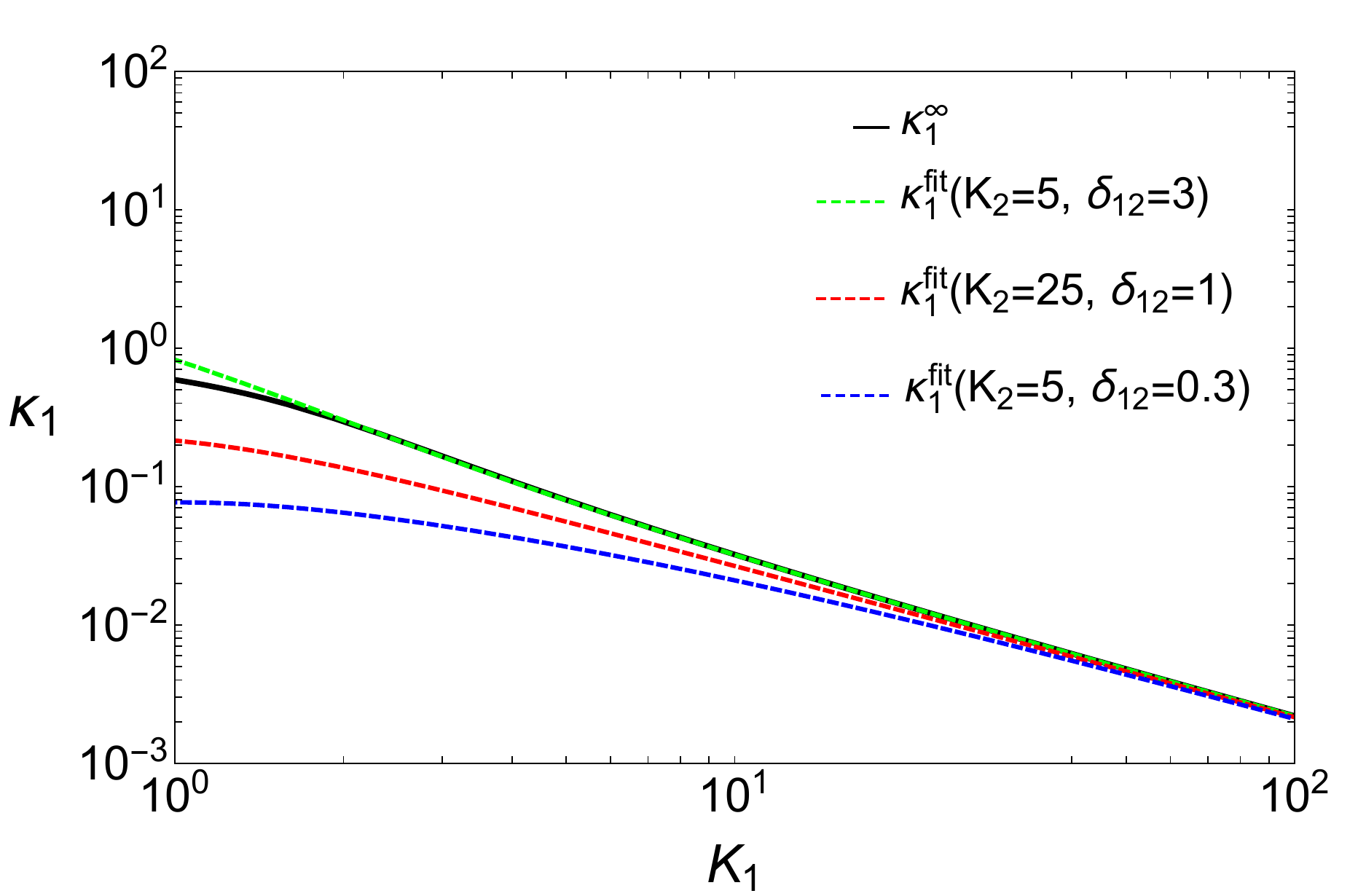}
\caption{Left: Possible error due to $\kappa_1^{\rm fit}$ as the efficiency factor in a hierarchical scenario, as a function of $\delta_{12}$. Right: Comparision of $\kappa_1^{\rm fit}$ to $\kappa_1^{\infty}$ for different values of $\delta_{12}$ and $K_2$. 
}\label{comp}
\end{figure}

It is also useful to have an expression for the efficiency factor for a strong washout scenario and any value of $\delta_{12}$. In this context, one can use \cite{Blanchet:2006dq}
\bea
\kappa_1^{\rm fit}=\frac{2K_1}{z_B\left(K_1+K_2^{(1-\delta_{12})^3}\right)\left(K_1+K_2^{1-\delta_{12}}\right)}\,,\label{fit}
\eea 
to scan the model, and estimate the minimum hierarchy of the RH neutrino masses for which $\kappa_1^{\rm fit}\rightarrow\kappa_1^\infty$. To quantify the goodness of this estimate, one can define an error function given by
\bea
{\rm Err}= \left|\frac{\kappa_1^{\rm fit}-\kappa_1^\infty}{\kappa_1^\infty}\right|\times 100\%\,.
\eea
In Fig.\,\ref{comp} (left panel), we show the error function for the two discussed cases, $\delta_{12}=1$ and $\delta_{12}=3$. It is obvious from this figure, that for the values of $\delta_{12}=1$ and $\delta_{12}=3$ chosen in Fig.\,\ref{f1}, the scope of error is always less that $\mathcal{O}(10\%)$.
In right panel of Fig.\,\ref{comp}, we show the comparison between $\kappa_1^{\rm fit}$ and $\kappa_1^{\infty}$ for the given values of $\delta_{12}$ in the strong washout regime. 
%Particularly the discussed cases ($\delta_{12}=3$, $K_{1,2}=5$; to be read from the green dashed line, $\delta_{12}=1$, $K_{1,2}=25$; to be read from the red dashed line). The blue dashed line is for $\delta_{12}=0.3$ which shows how one needs larger values for the decay parameter $K_1$ to circumvent the washout effect by $N_2$.
Clearly, if $\delta_{12}=0.3$ (blue dashed line), one needs larger values for the decay parameter $K_1$ to circumvent the washout effect by $N_2$. Therefore, Eq.\,\ref{fit} is also a reasonably good analytic approximation that can be used in the computation. \\
 
Thus, given the ranges of $K_1$ and $K_2$, one can do a random scanning over $\delta_{12}$ for each pair of $K_{1,2}$ to compare $\kappa_1(z)$ of Eq.\,\ref{effi1} or $\kappa_1^{\rm fit}$ of Eq.\,\ref{fit} to $\kappa_1^{\infty}$ upto desired accuracy, and extract the minimum values of $\delta_{12}$ needed to probe a perfectly valid hierarchical $N_1$DS. We shall show in the next section that lowering the value of $\delta_{12}$ has two major consequences. Firstly, for low values of $\delta_{12}$, one enhances the CP asymmetry parameter, which in turn increases the magnitude of the asymmetry due to an enhancement in the loop functions (particularly in the self energy contribution \cite{lep5}). Secondly, when flavour effects are accounted for, the contribution from $N_2$ \cite{Engelhard:2006yg,Blanchet:2011xq} to the final asymmetry plays an important role in a successful leptogenesis. Thus, given a particular flavour regime, lowering the value of $\delta_{12}$ enables us to extract information regarding $N_2$-leptogenesis over a wide range of RH neutrino mass scale. 
%%%%%%%%%%%%%%%%%%%
%%%%%%%%%%%%%%%%%%%
\section{Flavour effects and importance of $N_2$-leptogenesis}\label{s4}
%%%%%%%%%%%%%%%%%%%
%%%%%%%%%%%%%%%%%%%
The one flavour regime (1FR) is typically characterised by $M_i >10^{12}$ GeV where all the charged lepton flavours are out of equilibrium, and thus the lepton doublet $\ket{\ell_i}$ produced by the decay of the RH neutrinos can be written as a coherent superposition of the corresponding flavour states $\ket{\ell_{\alpha}}$ as,
\bea
\ket{\ell_i}&=&\mathcal{A}_{i\alpha} \ket{\ell_\alpha} \hspace{1cm} (i=1,2,3; \alpha=e,\mu,\tau)\\
\ket{\bar{\ell}_i}&=&\bar{\mathcal{A}}_{i\alpha} \ket{\bar{\ell}_\alpha} \hspace{1cm} (i=1,2,3; \alpha=e,\mu,\tau)\,,
\eea
where the amplitudes are given by
 \bea
 \mathcal{A}_{i\alpha}^0 =\frac{m_{D_{i\alpha}}}{\sqrt{(m_Dm_D^\dagger)_{ii}}}\hspace{1cm}{\rm and}\hspace{1cm}\bar{\mathcal{A}}_{i\alpha}^0 =\frac{m^*_{D_{i\alpha}}}{\sqrt{(m_Dm_D^\dagger)_{ii}}}.
 \eea
 Since there is hardly any interaction to break the coherence of the quantum states before it inversely decays to $N_1$, the asymmetry will be produced along the direction of $\ket{\ell_i}$(or $\ket{\bar{\ell}_i}$) in the flavour space. However, this is not the case if $M_i<10^{12}$ GeV, since below this scale, flavour effects become important. We give a brief overview of the flavour effects at play during leptogenesis in this section.
 
The flavour effects are taken into account by defining the branching ratios into individual flavours as $P_{i\alpha}=|\mathcal{A}_{i\alpha}|^2$ and $\bar{P}_{i\alpha}=|\bar{\mathcal{A}}_{i\alpha}|^2$. As a result, the decays into individual flavours could be written as $\Gamma_{i\alpha}\equiv P_{i\alpha}$ $\Gamma_i$ and $\bar{\Gamma}_{i\alpha}\equiv\bar{P}_{i\alpha}\bar{\Gamma}_i$ with $\sum_{\alpha}(P_{i\alpha},\bar{P}_{i\alpha})=1$. It is also convenient to introduce the flavoured decay parameter $K_{i\alpha}$ given by 
 \bea
 K_{i\alpha}=\frac{\Gamma_{i\alpha}+\bar{\Gamma}_{i\alpha}}{H(T=M_i)}\simeq \frac{ P_{i\alpha}^0(\Gamma_i+\bar{\Gamma}_i)}{H(T=M_i)}\equiv P_{i\alpha}^0 K_i\equiv\frac{|m_{D_{i\alpha}}|^2}{M_i m^*},\label{fldecay}
 \eea
 where $m^*\simeq 10^{-3}\,{\rm eV}$ is the equilibrium neutrino mass. These flavoured probabilities can be re-written as 
 \bea
 P_{i\alpha}&=& P_{i\alpha}^0+\frac{\Delta P_{i\alpha}}{2}\,,\\ 
 \bar{P}_{i\alpha}&=& P_{i\alpha}^0-\frac{\Delta P_{i\alpha}}{2}\,,
 \eea
 where 
 \bea
 P_{i\alpha}^0&=&\frac{1}{2}\left(P_{i\alpha}+\bar{P}_{i\alpha}\right)\,,\\
 \Delta P_{i\alpha}&=&P_{i\alpha}-\bar{P}_{i\alpha}\,
 \eea
 are the tree level projectors. Here $\Delta P_{i\alpha}$, the difference between the tree level and the loop level projectors, arises from the fact that $\mathcal{A}_{i\alpha}\neq \bar{\mathcal{A}}_{i\alpha}$ \cite{lep3}, except at tree level. This allows us to define the flavoured CP asymmetry parameter $\varepsilon_{i\alpha}$ (see Eq.\,\ref{epsi}) as 
 \bea
 \varepsilon_{i\alpha}=P_{i\alpha}^0\varepsilon_i+\Delta P_{i\alpha}/2\,.\label{pflp}
 \eea
 Thus, due to the incorporation of flavour effects, an extra amount of CP violation, characterised by $\Delta P_{i\alpha}$, is generated. Note that in Eq.\ref{pflp} one can have CP violation in each flavour even if the total CP asymmetry is vanishing \cite{pfl}. Typically, the effect of $\Delta P_{i\alpha}$ can be neglected in the washout terms, however, this is not the case for $\varepsilon_{i\alpha}$.

In the regime $10^{9}$ GeV $<M_i<10^{12}$ GeV, the $\tau$ flavored lepton comes into equilibrium, thereby breaking the coherent evolution of $\ket{\ell_i}$ before it inverse decays to $N_i$. As a result, $\ket{\ell_i}$ is projected onto a two flavour basis, characterised by the eigenstates along the directions of $\tau$, and perpendicular to it $(\tau_{i}^\perp)$, which is essentially a coherent superposition of the $\mu$ and the $e$ flavour. In the three flavour regime, i.e. all $M_i<10^{9}$ GeV, the $\mu$ lepton also comes into equilibrium, thus breaking the coherent evolution of the states along $\tau_{i}^\perp$. This allows for the individual resolution of all the flavours. Thus, calculating the asymmetry produced requires tracking the lepton asymmetry in the relevant flavours.

 For example, in the 2FR, the lepton asymmetry has to be tracked in $\tau$ and $\tau_i^\perp$. The Boltzmann equations can be written as
\bea
\frac{dN_{N_i}}{dz}&=&-D_i(N_{N_i}-N_{N_i}^{\rm eq}), ~{\rm with}~i=1,2. \label{be1}\\
\frac{dN_{{\Delta}_{\alpha}}}{dz}&=&-\sum_{i=1}^2\varepsilon_{i\alpha} D_i(N_{N_i}-N_{N_i}^{\rm eq})-\sum_{i=1}^2P_{i\alpha}^0W_i^{\rm ID}N_{{\Delta}_{\alpha}}\,.\label{be2}
\eea
The asymmetry in the flavour $\alpha$ is given by
\bea
N_{\Delta_{\alpha}}=-\sum_{i}^2\varepsilon_{i\alpha}\kappa_{i\alpha}\,.
\label{fep}
\eea
with the efficiency factor
\bea
\kappa_{i\alpha} (z)=-\int_{z_{\rm in}}^\infty \frac{dN_{N_i}}{dz^\prime}e^{-\sum_{j}\int_{z^\prime}^zP_{j\alpha}^0 W_{j}^{\rm ID}(z^{\prime\prime})dz^{\prime\prime}}dz^\prime\,.\label{effi}
\eea
With this definition, the final baryon to photon ratio is 
\bea
\eta_{B}=0.96\times10^{-2}\sum_{\alpha}N_{\Delta_{\alpha}}\,.
\eea

In the hierarchical limit of the RH neutrino masses, Eq.\,\ref{fep} can be simplified as 
\bea
N_{\Delta_{\alpha}}=-\varepsilon_{1\alpha}\kappa_{1\alpha}^\infty-\varepsilon_{2\alpha}\kappa_{2 \alpha}^\infty e^{-3\pi K_{1\alpha}/8}\,,\label{fep1}
\eea
where the first term is the asymmetry generated by $N_1$, and the second term is the asymmetry generated by $N_2$, subjected to $N_1$-washout.
However, there are two important issues, which are usually overlooked in the leptogenesis studies of models with flavour symmetries. 

 i) A pure $N_1$-leptogenesis scenario which is studied in most of the neutrino mass models, requires large values of the $N_1$ decay parameter $K_{1\alpha}$ to washout the contribution from $N_2$. Thus given a neutrino mass model constrained by 3$\sigma$ oscillation data, one has to check the strength of $K_{1\alpha}$ so that the second term of Eq.\ref{fep1} can be neglected.
 
\begin{figure}[!t]
\begin{center}
\includegraphics[scale=0.5]{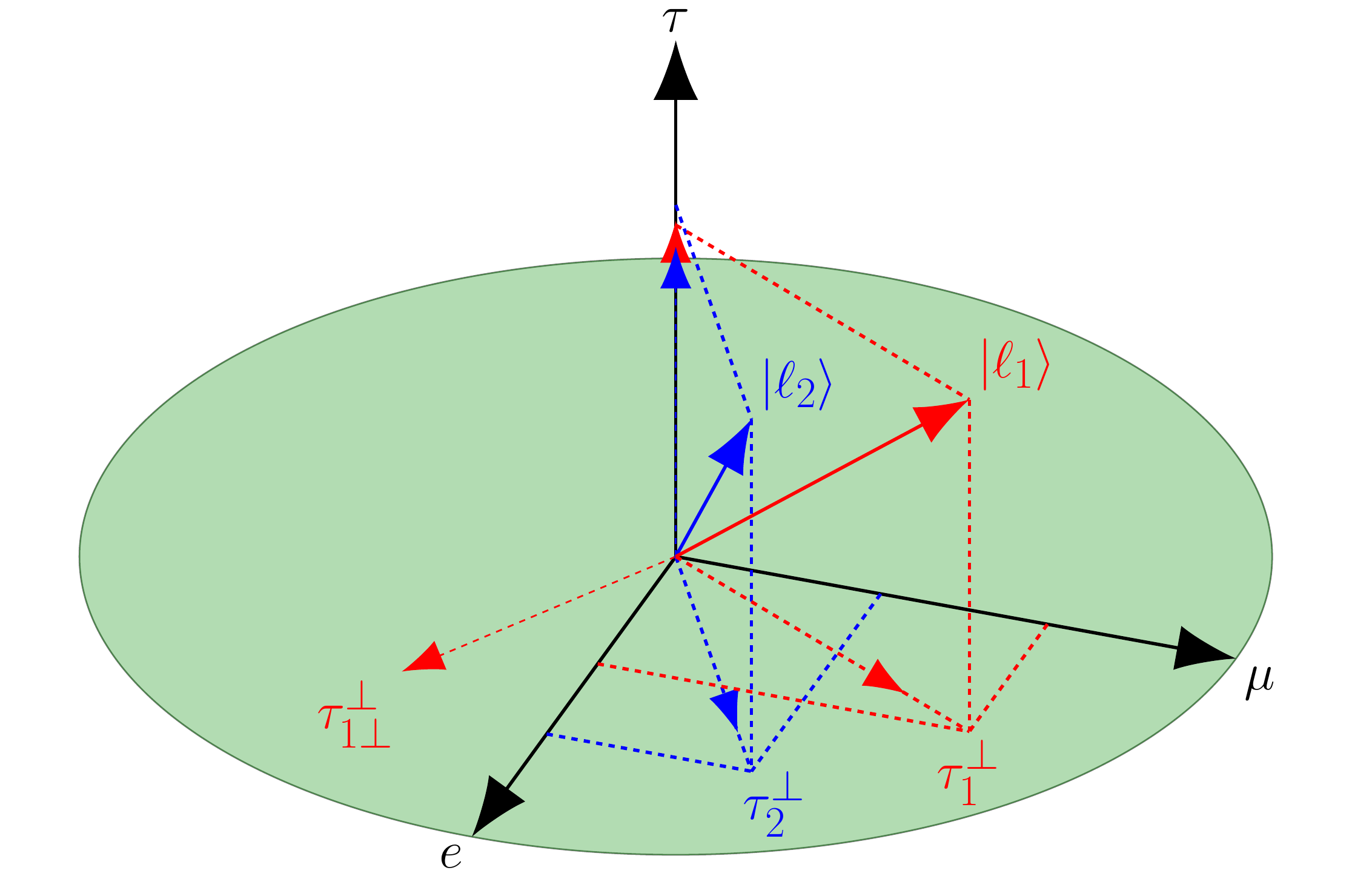}
\caption{Ilustration of the two flavour regime for two RH neutrino model.}\label{axes}
\end{center}
\end{figure} 
 
 ii) Most importantly, if the masses of both the RH neutrinos are in the 2FR, i.e., $10^{9}$ GeV $<M_i<10^{12}$ GeV, after the $\tau$-interactions of both the states $\ket{\ell_1}$ and $\ket{\ell_2}$, the resultant states orthogonal to the $\tau$ flavour will not be in the same direction on the $e-\mu$ plane. This is demonstrated in Fig.\,\ref{axes}, where the new directions are denoted by $\tau_1^\perp$ and $\tau_2^\perp$ respectively. This is simply due the fact that, in general $\mathcal{A}_{1\alpha}\neq \mathcal{A}_{2\alpha}$, and hence, there is no reason for the states to maintain a common direction.

 Henceforth, we denote the $\tau_i^\perp$ states as $\ket{\ell_1^{\tau^\perp}}$ and $\ket{\ell_2^{\tau^\perp}}$, which are given by
\bea
\ket{\ell_1^{\tau^\perp}}=\frac{\mathcal{A}_{1e}}{\sqrt{|\mathcal{A}_{1e}|^2+|\mathcal{A}_{1\mu}|^2}}\ket{\ell_e}+\frac{\mathcal{A}_{1\mu}}{\sqrt{|\mathcal{A}_{1e}|^2+|\mathcal{A}_{1\mu}|^2}}\ket{\ell_\mu},\\
\ket{\ell_2^{\tau^\perp}}=\frac{\mathcal{A}_{2e}}{\sqrt{|\mathcal{A}_{2e}|^2+|\mathcal{A}_{2\mu}|^2}}\ket{\ell_e}+\frac{\mathcal{A}_{2\mu}}{\sqrt{|\mathcal{A}_{2e}|^2+|\mathcal{A}_{2\mu}|^2}}\ket{\ell_\mu}.
\eea
In order to guess how much asymmetry generated by $N_2$ along $\tau_2^\perp$ can be washed out by the interactions between the Higgs and the component of $\ket{\ell_2^{\tau^\perp}}$ along $\ket{\ell_1^{\tau^\perp}}$ ($N_1$ inverse decay), one has to calculate the probability of $\ket{\ell_2^{\tau^\perp}}$ being in the $\ket{\ell_1^{\tau^\perp}}$ state . Note that for the $N_1$ inverse decay, only $\left(\braket{\ell_1^{\tau^\perp}|\ell_2^{\tau^\perp}}\right)\ket{\ell_1^{\tau^\perp}}$ will interact with the Higgs, whereas $\left(\braket{\ell_{1\perp}^{\tau^\perp}|\ell_2^{\tau^\perp}}\right)\ket{\ell_{1\perp}^{\tau^\perp}}$, which is perpendicular to $\ket{\ell_1^{\tau^\perp}}$, will be blind to it. Thus, the asymmetry in the direction of $\ket{\ell_{1\perp}^{\tau^\perp}}$ will escape the $N_1$ washout and survive as a pure contribution from $N_2$.

The overlap probability $p_{12}$ can be calculated as 
\bea
p_{12}\equiv |\braket{\ell_1^{\tau^\perp}|\ell_2^{\tau^\perp}}|^2=\frac{K_1K_2}{K_{1\tau^\perp} K_{2\tau^\perp}}\frac{|(m_D^*)_{1e}(m_D)_{2e}+(m_D^*)_{1\mu}(m_D)_{2\mu}|^2}{h_{11}h_{22}}\,,\label{p12}
\eea
where $h_{ii}=(m_Dm_D^\dagger)_{ii}$. 

With this understanding, the RHS of Eq.\,\ref{fep1} can be split into three parts
\bea
N_{\Delta_{\tau}}&=&-\varepsilon_{1\tau}\kappa_{1\tau}^\infty-\varepsilon_{2\tau}\kappa_{2 \tau}^\infty e^{-3\pi K_{1\tau}/8},\label{fep2}\\
N_{\Delta_{\tau_1^\perp}}&=&-\varepsilon_{1\tau^\perp}\kappa_{1\tau^\perp}^\infty-p_{12}\varepsilon_{2\tau^\perp}\kappa_{2 \tau^\perp}^\infty e^{-3\pi K_{1\tau^\perp}/8},\label{fep3}\\
N_{\Delta_{\tau_{1\perp}^\perp}}&=&-(1-p_{12})\varepsilon_{2\tau^\perp}\kappa_{2 \tau^\perp}^\infty\,, \label{fep4}
\eea 
where the final $B-L$ asymmetry is given by
\bea
N_{B-L}^f=N_{\Delta_{\tau}}+N_{\Delta_{\tau_1^\perp}}+N_{\Delta_{\tau_{1\perp}^\perp}}\,.
\eea

Note that in a situation where a strong washout by the $N_1$ inverse decay prevails, the second term in Eq.\,\ref{fep2} and \ref{fep3} can be dropped. Hence, the $p_{12}\rightarrow 1$ would imply a pure $N_1$-leptogenesis. In the literature, along with a strong $N_1$-washout, it is usually assumed that $p_{12}=1$, which is not true in general. \\

 Another interesting situation arises when $M_2$ is in the two flavour regime and $M_1$ is in the three flavour regime. In this case, the produced asymmetry by $N_2$ in two flavour regime will be washed out by $N_1$ in the three flavour regime. Therefore, at the end of $N_1$-washout, we need to track the final asymmetry in individual flavours ($e,\mu,\tau$). Thus, the asymmetry in each flavour can be written as
\bea
N_{\Delta_{\tau}}&=&-\varepsilon_{1\tau}\kappa_{1\tau}^\infty-\varepsilon_{2\tau}\kappa_{2 \tau}^\infty e^{-3\pi K_{1\tau}/8},\label{fep5}\\
N_{\Delta_{\mu}}&=&-\varepsilon_{1\mu}\kappa_{1\mu}^\infty-\frac{K_{2\mu}}{K_{2\tau^\perp}}\varepsilon_{2\tau^\perp}\kappa_{2 \tau^\perp}^\infty e^{-3\pi K_{1\mu}/8},\label{fep6}\\
N_{\Delta_{e}}&=&-\varepsilon_{1e}\kappa_{1e}^\infty-\frac{K_{2e}}{K_{2\tau^\perp}}\varepsilon_{2\tau^\perp}\kappa_{2 \tau^\perp}^\infty e^{-3\pi K_{1e}/8},\label{fep7}
\eea
where the final $B-L$ asymmetry now is given by
\bea
N_{B-L}=\sum_{\alpha}N_{\Delta_{\alpha}}\hspace{1.5cm}(\alpha=e,\mu,\tau)\,.
\eea

\begin{figure}[!t]
\begin{center}
 \includegraphics[scale=0.69]{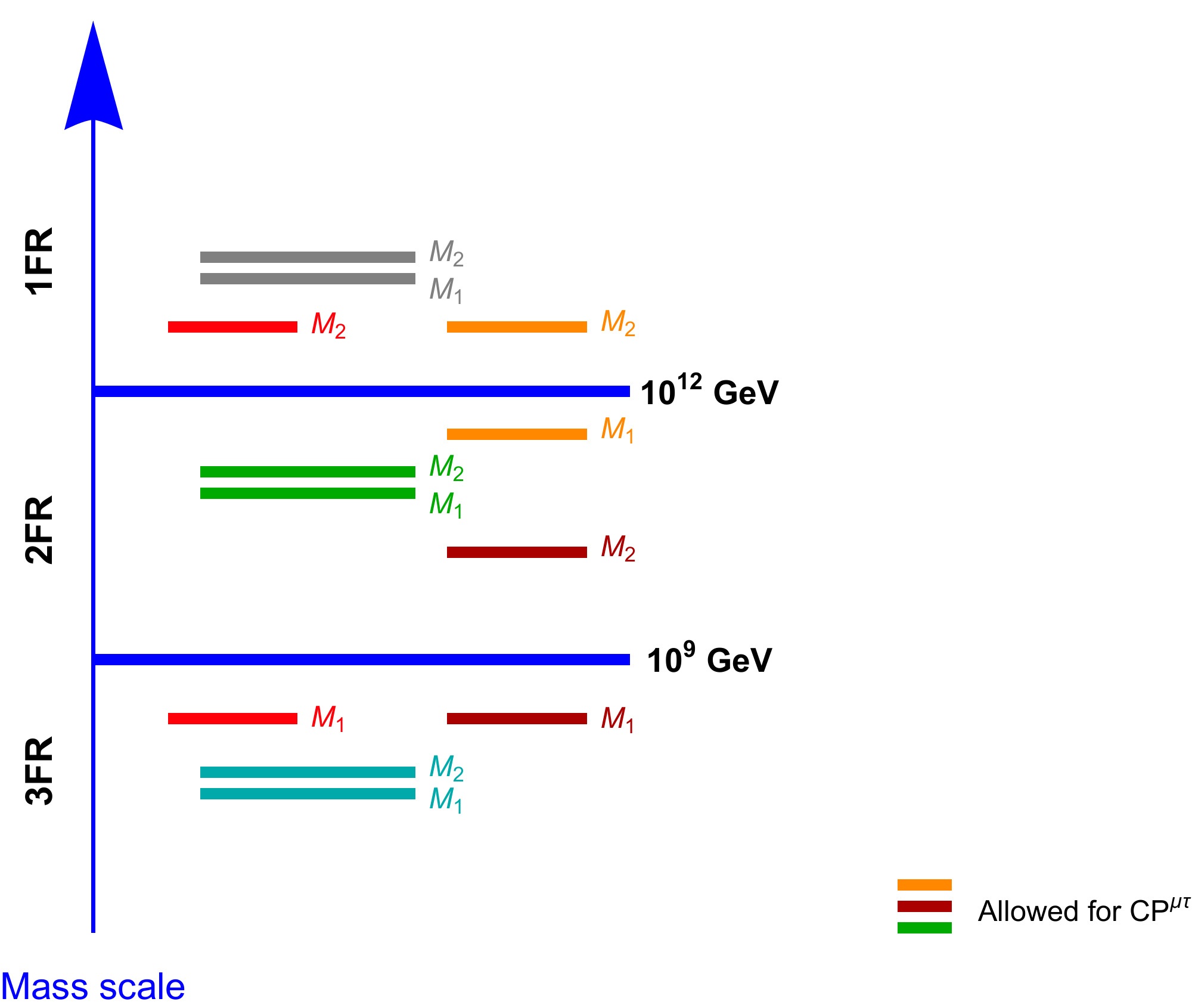} 
 \caption{Various mass pattern in a leptogenesis scenario dominated by two right handed neutrinos. A particular RH neutrino mass which is either above $10^{12}$ GeV or below $10^{9}$ GeV,  can not generate baryon asymmetry in the $\rm CP^{\mu\tau}$ framework.}\label{fig7}
\end{center}
\end{figure}

Note that in Eq.\,\ref{fep5}--\ref{fep7}, the first term is the contribution to the final asymmetry from $N_1$ which produces the lepton asymmetry in 3FR, where one can distinguish each of the three flavours. There could be other possibilities such as $M_i<10^9$ GeV, $M_i>10^{12}$, and $M_2>10^{12}$ GeV but $M_1<10^{9}$ GeV as shown in Fig.\,\ref{fig7}. Among these three possibilities, whilst the first one is not compatible to the standard thermal hierarchical leptogenesis scenario due to Davidson-Ibarra bound on $M_i$ \cite{ibarra}, for the rest of the cases, successful leptogenesis cannot be realized unless we invoke some special conditions.

%%%%%%%%%%%%%%%%
%%%%%%%%%%%%%%%%
\section{Leptogenesis in the $\rm CP^{\mu\tau}$ symmetric model}\label{s5}
\begin{figure}[!t]
\begin{center}
\includegraphics[scale=0.45]{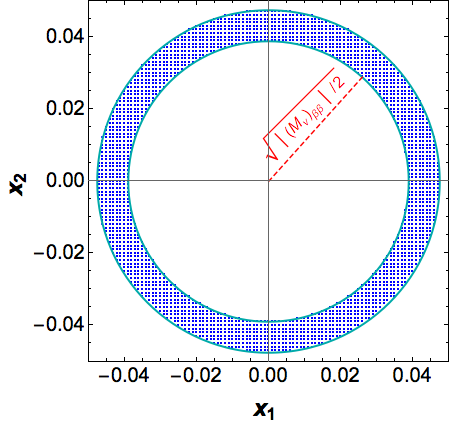} \includegraphics[scale=0.45]{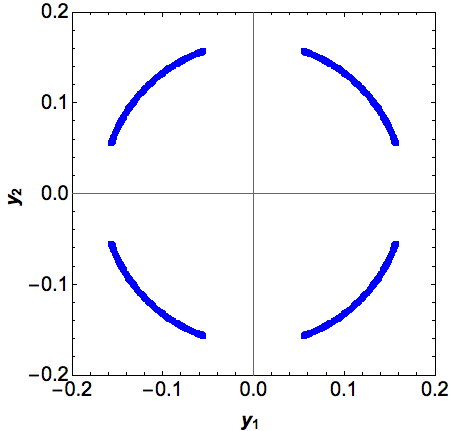}
\caption{Parameter space of ${\rm CP^{ \mu\tau}}$ symmetric mass matrix within a two RH neutrino scenario:  $x_1$ vs $x_2$ (left) and $y_1$ vs $y_2$ (right), where these dimensional parameters (in $ \sqrt{\rm eV}$) are defined in Eq.\ref{primed}. }\label{fig5}
\end{center}
\end{figure}

%%%%%%%%%%%%%%%%
%%%%%%%%%%%%%%%%
To carry out a numerical computation pertaining to a successful leptogenesis, we need to constrain the model parameters of Eq.\,\ref{mnuseesaw} with the present neutrino oscillation data \cite{globalfit}. For a normal neutrino mass ordering with solar and atmospheric mass squared differences, $\Delta m_{12}^2=7.39^{+0.21}_{-0.20}\times10^{-5}{\rm eV}^2$ and $\Delta m_{31}^2=2.52^{+0.033}_{-0.032}\times10^{-3}{\rm eV}^2$, the current global-fit values of the three mixing angle and the Dirac CP phases are tabulated in Table \ref{t1}. To this end, we follow the exact diagonalization procedure of a $3\times 3$ light neutrino mass matrix, first demonstrated in \cite{Adhikary:2013bma}. This gives $-150^\circ < \theta < 150^\circ$, while the ranges of the other parameters are shown in the Fig.\,\ref{fig5}. 

The shape of the allowed parameter space in Fig.\,\ref{fig5} could intuitively be inferred as follows. For a fixed value of $|(M_\nu)_{ee}|$ or $|(M_\nu)_{\mu\tau}|$ (say $c$), the solution is that of a circle\footnote{Though it has been noticed that vanishing or close to vanishing values of $y_{1,2}$ are not compatible with present neutrino oscillation data. Discussion regarding the parameter sapce can be found in Ref.\cite{cp16}.}, given by $x_1^2+x_2^2=c$ or $y_1^2+y_2^2=c$. Considering the left panel of Fig.\,\ref{fig5}., since the radii of each of these circles are related to the neutrinoless double-beta decay parameter $\sqrt{|(M_\nu)_{\beta\beta}|/2}$ (cf. Eq.\,\ref{mnuseesaw}), there exists an upper limit $\sim$ 5 meV and a lower limit $\sim$ 3 meV (represented by the cyan circles) on $|(M_\nu)_{\beta\beta}|$. However both the limits on $|(M_\nu)_{\beta\beta}|$ are beyond the sensitivity reach of the present experiments such as GERDA \cite{gerda}, KamLAND-Zen \cite{kzen}, EXO \cite{exo} etc., as well as the next generation experiments\cite{Agostini:2017jim} like KamLAND2-Zen \cite{k2zen}, nEXO \cite{nexo}, CUPID \cite{cupid}, CUORE\cite{cuore}, LEGEND-1k \cite{l1k}. Thus, this model lacks testability from these experiments.
%\ms{Why discontinuity in right panel?}

\begin{figure}[!t]
\begin{center}
\includegraphics[scale=0.43]{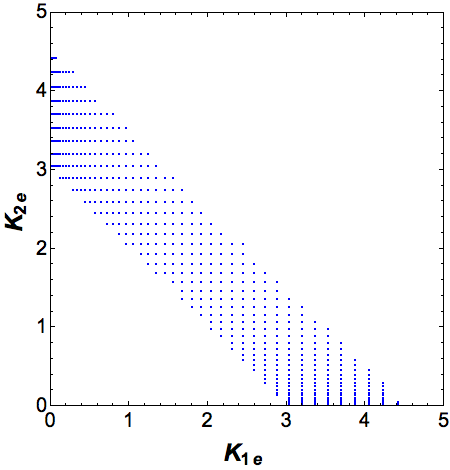}\hspace{6mm} \includegraphics[scale=0.45]{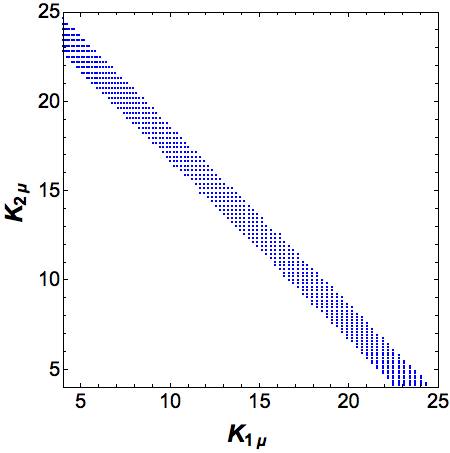}
\caption{Flavoured decay parameters for both the RH neutrinos.}\label{fig6}
\end{center}
\end{figure}

\begin{table}[!t]
\caption{Best-fit, 1$\sigma$ and 3$\sigma$ ranges of three mixing angles and the Dirac CP phase $\delta$ for NMO (\href{http://www.nu-fit.org/?q=node/12}{NuFIT\cite{globalfit}})}\label{t1}
\hspace{2cm}
\begin{tabular}{|c|c|}
\hline
&$\theta_{12}/^\circ$~~~~~~~~~~~$\theta_{23}/^\circ$~~~~~~~~~~~$\theta_{13}/^\circ$~~~~~~~~~~~$\delta/^\circ$\\
\hline
\hline
${\rm bf}\pm1\sigma$&$33.82^{+0.78}_{-0.76}$~~~~~~~$49.6^{+1.0}_{-1.2}$~~~~~~~$8.61^{+0.13}_{-0.13}$~~~~~~~$215^{+40}_{-29}$\\
\hline
$3\sigma$&$31.61\rightarrow 36.27$~~~$40.3\rightarrow 52.4$~~~$8.22\rightarrow 8.99$~~~$125\rightarrow 392$\\
\hline
\end{tabular}
\end{table}
From Eq.\,\ref{mnuseesaw} and Eq.\,\ref{fldecay}, it is trivial to derive analytic correlations between the flavoured decay parameters as
\bea
K_{2e}&=&\frac{|(M_\nu)_{\beta\beta}|}{m^*}-K_{1e}\,,\\
K_{2\mu}&=&\frac{|(M_\nu)_{\mu\tau}|}{m^*}-K_{1\mu}\,
\eea
which are shown in Fig.\,\ref{fig6}. There are two interesting observations to be made from these plots. Firstly, note that the decay parameters in the electron flavour can have approximately vanishing values, as is clear from the left panel. Secondly, the decay parameters in the muon flavour or tau flavour (in this case $K_{i\mu}=K_{i\tau}$) have a lower bound ($\sim 5$) due to the discontinuity in parameter space of $y_1$ and $y_2$ (see right panel of Fig.\,\ref{fig5}). We see later that these ranges of the decay parameters have very interesting consequences on the process of leptogenesis in this model.

 Let us first discuss two interesting mass patterns of the RH neutrinos: $M_i>10^{12}$ GeV and $M_i<10^9$ GeV \footnote{Both these mass patterns have been discussed in literature, e.g., for the first one see \cite{cp3,cplepto1} and for the seconed one, see \cite{Samanta:2018efa}. We recall the discussion here for comprehensiveness. } . 
 First of all, for the one flavour regime ($M_i>10^{12}$ GeV), the second term in Eq.\,\ref{ncp} vanishes when summed over `$\alpha$', i.e, ${\rm Im}\{{h}_{ji}({m_D})_{i\alpha} (m_D^*)_{j\alpha}\}={\rm Im}[|h_{ji}|^2]=0$. The first term, however, is proportional to ${\rm Im}\{{h}_{ij}^2\}$. Using Eq.\,\ref{mdseesaw}, one can show that $h=m_Dm_D^\dagger$ is a real matrix \cite{cp16}. Thus, the flavour-summed CP asymmetry $\varepsilon_i=\sum_{\alpha}\varepsilon_{i\alpha}$ vanishes for any $i$. Therefore, successful leptogenesis is not possible in the unflavoured regime. Interestingly, $\varepsilon_{ie}$ is also vanishing, since the phases associated with the relevant parameters of $m_D$ will cancel when one uses Eq.\,\ref{ncp} to calculate the CP asymmetry in the electron flavour. Thus in this model, $\varepsilon_{i\mu}\equiv\Delta P_{i\mu}/2=-\varepsilon_{i\tau}$. On the other hand, if all the RH neutrino masses are in the three flavour regime $M_i<10^{9}$ GeV, one might wonder whether there would be possibilities for a resonant leptogenesis \cite{lep5,Deppisch:2010fr}. However, in \cite{Samanta:2018efa}, it has been analytically argued that due to the typical structure of the symmetry (the efficiency factors in $\mu$ and $\tau$ flavour are same), such a possibility still leads to a vanishing asymmetry even after taking into account the flavour coupling effects \cite{fl2,Barbieri:1999ma,Antusch:2010ms}.
 
 Another interesting possibility is to consider $M_2>10^{12}$ GeV, and $M_1<10^{12}$ GeV. In that case, since the asymmetry is produced by $N_2$ in the unflavoured regime and $\varepsilon_i=\sum_{\alpha}\varepsilon_{i\alpha}=0$, the final baryon asymmetry only has contributions from $N_1$ \footnote{$N_2$ might contribute to the final asymmetry via phantom terms\cite{Blanchet:2011xq}. However, phantom leptogenesis in this context is beyond the scope of this study.}. As a result, all the results derived in Ref. \cite{cp3,cp16} will be valid upto minor changes due to the newly released global-fit data \cite{globalfit}. 
 
In this paper, we shall focus on the following mass patterns: i) $10^9$ GeV $<M_{1,2}<10^{12}$ GeV, and ii) $10^9$ GeV $<M_{2}<10^{12}$ and $M_1<10^9$ GeV. Before discussing these cases explicitly, we list the flavoured CP asymmetry parameters in this model. Using Eq.\,\ref{mdseesaw} and Eq.\,\ref{ncp} the $\varepsilon_{i\alpha}$ can be obtained as 
\bea
\varepsilon_{ie}=0,\hspace{1mm}\varepsilon_{i\mu}=-\xi_i\frac{g^\prime(x_{ij})}{4\pi v^2}\left[\frac{(a_ia_j+b_ib_j\cos\theta)b_ib_j\sin\theta}{a_i^2+b_i^2}\right]=-\varepsilon_{i\tau}\,, \quad i\neq j(=1,2),\label{CPmod}
\eea
where $g^\prime(x_{ij})$ is given by
\bea
g^\prime(x_{ij})\simeq[f(x_{ij})+\sqrt{x_{ij}}/(1-x_{ij})]+(1-x_{ij})^{-1}\equiv g_1(x_{ij})+g_2(x_{ij})\,,\label{gpr}
\eea
and $\xi_i=\pm 1$ for $i=1$ and 2 respectively. Using Eq.\,\ref{primed} we can now simplify Eq.\,\ref{CPmod} for $i=1$ as 
\bea
\varepsilon_1^\mu=-\frac{g^\prime(x_{12})M_2}{4\pi v^2}\left[\frac{(x_1x_2+y_1y_2\cos\theta)y_1y_2\sin\theta}{x_1^2+y_1^2}\right]=-\varepsilon_1^\tau\,, \label{epn1}
\eea 
which in the strong hierarchical limit can further be simplified as 
\bea
\varepsilon_1^\mu\simeq\frac{3M_1}{8\pi v^2}\left[\frac{(x_1x_2+y_1y_2\cos\theta)y_1y_2\sin\theta}{x_1^2+y_1^2}\right]=-\varepsilon_1^\tau\,. \label{epn1s}
\eea 
Similarly for $i=2$ the CP asymmetry parameter can be calculated as
\bea
\varepsilon_2^\mu=\frac{g^\prime(x_{21})M_1}{4\pi v^2}\left[\frac{(x_1x_2+y_1y_2\cos\theta)y_1y_2\sin\theta}{x_2^2+y_2^2}\right]=-\varepsilon_2^\tau\,. \label{epn2}
\eea 
Armed with these equations, we can proceed toward a systematic discussion of leptogenesis for the relevant cases.

\subsection{Two flavour regime: $10^9 \,{\rm GeV} <M_{1,2}<10^{12} \,{\rm GeV}$}

The $N_1$-decay parameters in the muon and tau flavour are strong enough \cite{lep6,lep3} to washout any pre-existing asymmetry (see right panel of Fig.\,\ref{fig6}). Thus, all the terms which contain the exponential washout factors in Eq.\,\ref{fep2} and Eq.\,\ref{fep3} can be neglected. Therefore the total $N_{B-L}$ asymmetry can be written as 
\bea
N_{B-L}&=&-(\varepsilon_{1\tau}\kappa_{1\tau}^\infty+\varepsilon_{1\tau^\perp}\kappa_{1\tau^\perp}^\infty)-(1-p_{12}^\perp)\varepsilon_{2\tau^\perp}\kappa_{2\tau^\perp}^\infty\nonumber\\
&=&-\varepsilon_{1\tau}(\kappa_{1\tau}^\infty-\kappa_{1\tau^\perp}^\infty)-(1-p_{12}^\perp)\varepsilon_{2\mu}\kappa_{2\mu}^\infty\,,
\label{flaI}
\eea
where we use the fact, that $\varepsilon_{ie}=0$, $\varepsilon_{i\mu}=-\varepsilon_{i\tau}$, and the electron decay parameters are much weaker than the muon decay parameters. Clearly, the first term in Eq.\,\ref{flaI}, which is a contribution from $N_1$, is non-vanishing when $\kappa_{1\tau}^\infty\neq\kappa_{1\tau^\perp}^\infty$, i.e, when there is an asymmetric washout in the $\tau$ and $\tau^\perp$ flavour. The second term, driven by the muon flavour, is a pure contribution from $N_2$, and is non-zero when $p_{12}^\perp\neq1$. Using Eq.\,\ref{p12}, one can arrive at an expression for the probability $p_{12}^\perp$ as, 
\bea
p_{12}^\perp=\frac{4x_1^2x_2^2+y_1^2y_2^2+4x_1x_2y_1y_2 \cos\theta}{(2x_1^2+y_1^2)(2x_2^2+y_2^2)}\,.
\eea

\begin{figure}[!t]
\begin{center}
\includegraphics[scale=0.5]{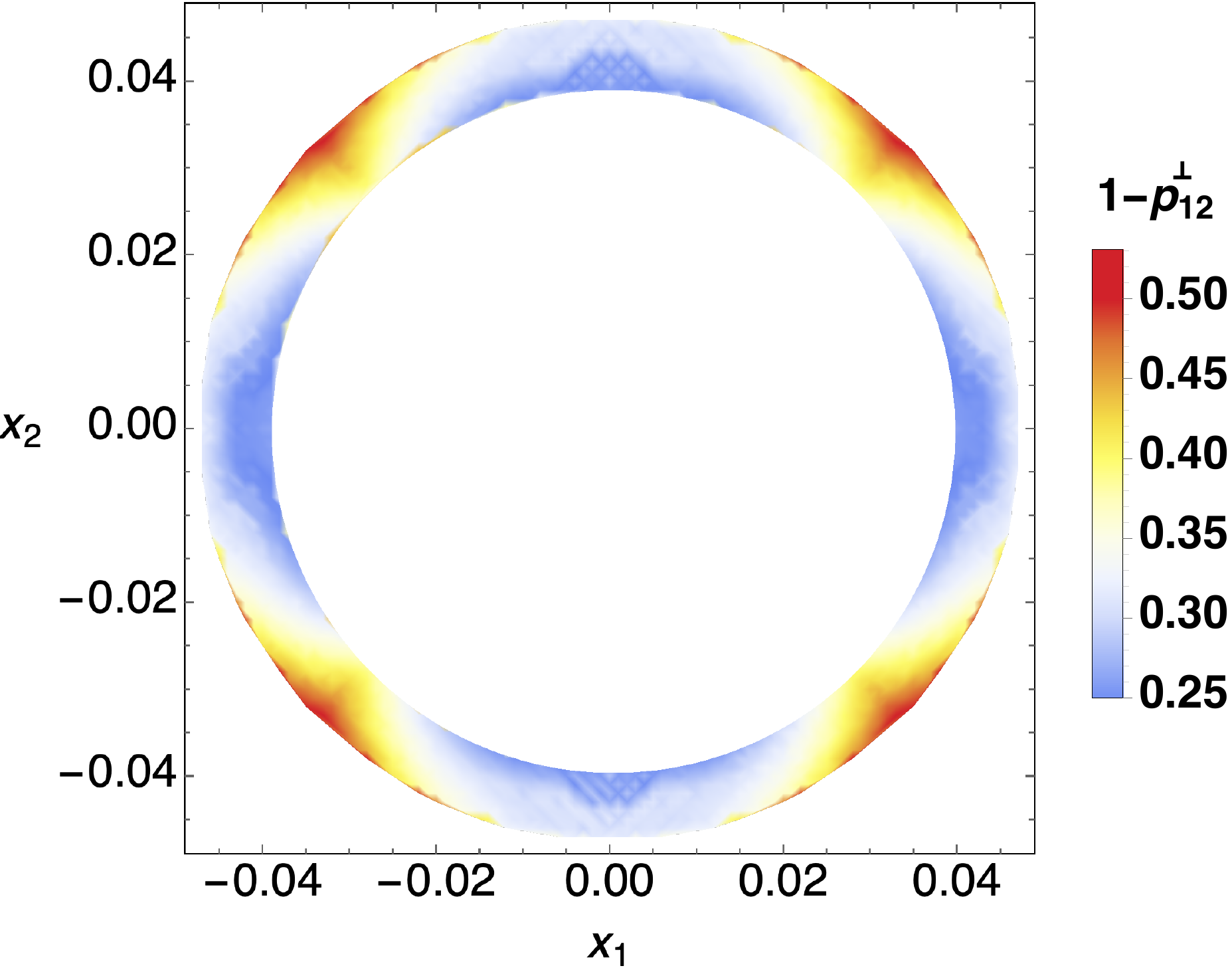}
 \caption{Plot showing the quantity $(1-p_{12}^\perp)$ with the model parameter $x_1$ and $x_2$. Since lepton asymmetry generated by $N_2$ is proportional to $(1-p_{12}^\perp)$, and clearly this never vanishes in this model, a pure $N_1$ dominated scenario is not possible. Note that $x_{1,2}$ are dimensional quantities, and are plotted in units of $ \sqrt{\rm eV}$. } \label{p12p}
\end{center}
\end{figure}

 In Fig.\,\ref{p12p} we show the variation of $(1-p_{12}^\perp)$ with the model parameter $x_1$ and $x_2$. An interesting fact is that $(1-p_{12}^\perp)$ never vanishes in this model. This means $N_2$ always contributes to the final asymmetry. In addition, one has a strong concentration of points towards the higher values ($\sim 0.5$) of $(1-p_{12}^\perp)$ which indicates there could be sizeable number of data points for which $N_2$ domination could be realized. In fact we show as we proceed, $N_2$ domination in this model is possible for a significant amount of parameter space ($\sim$ 26$\%$).

\begin{figure}[!t]
\includegraphics[width=3.5in, height=2.2in]{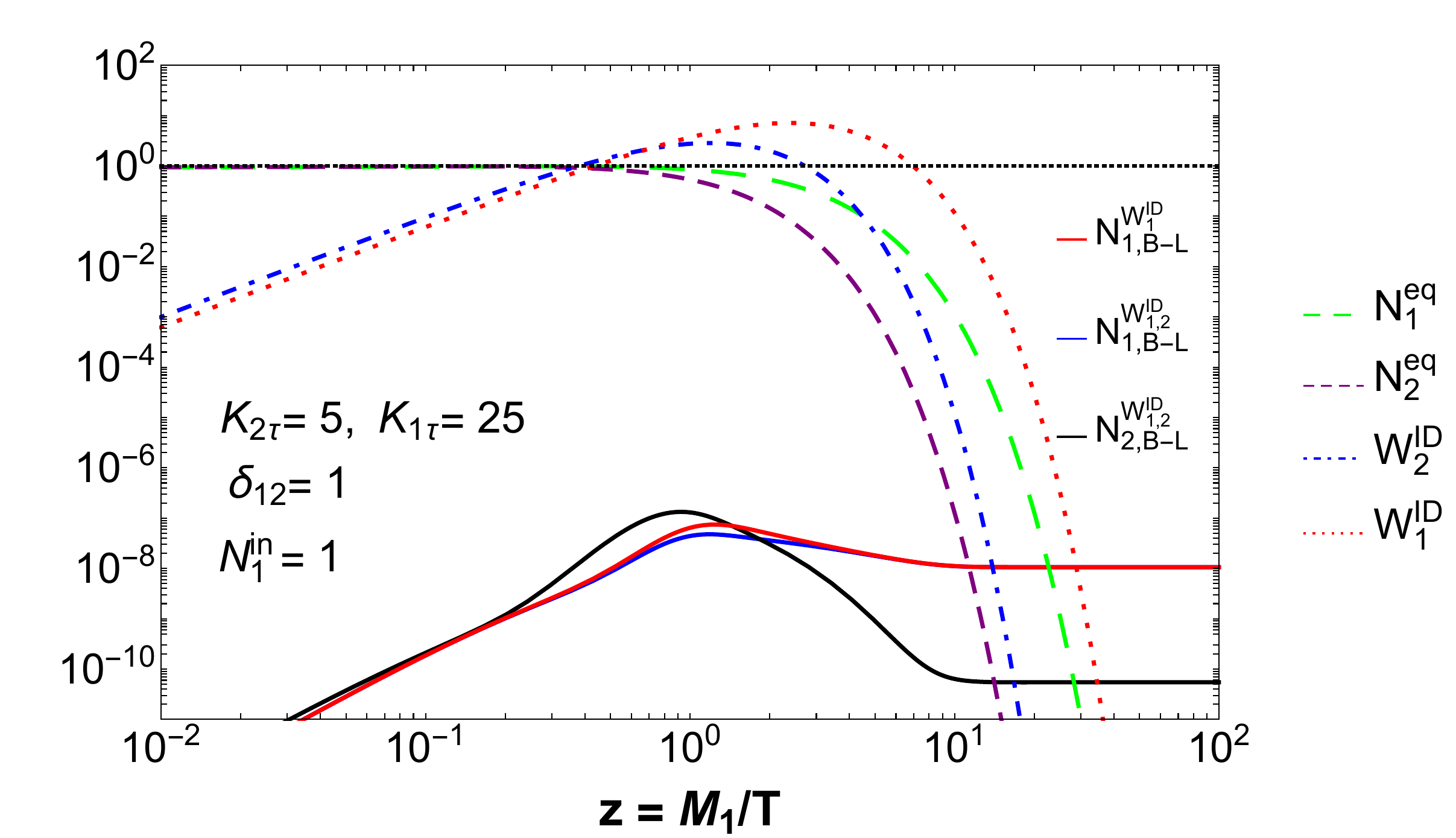}~\includegraphics[width=3.5in, height=2.2in]{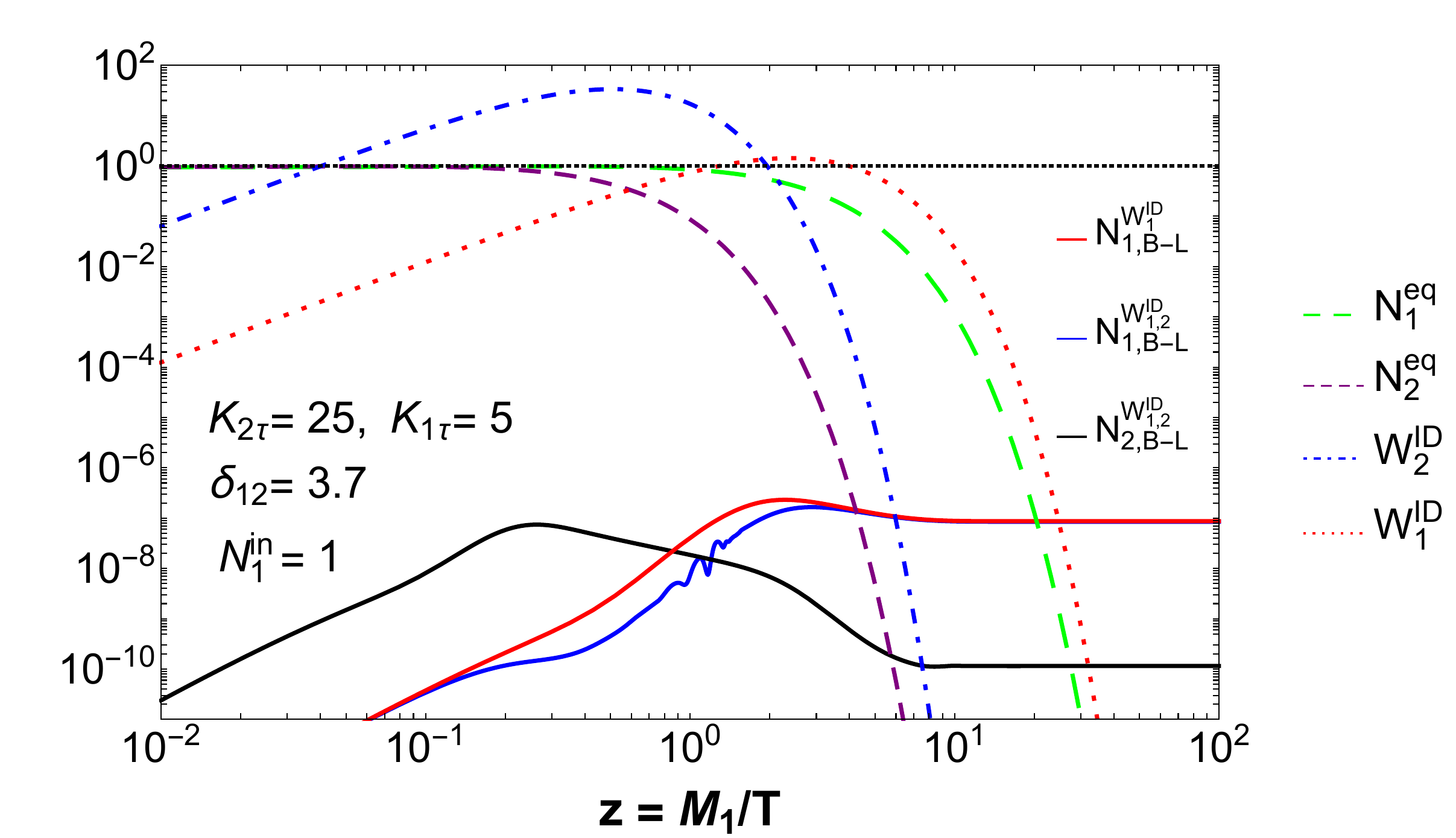}\\
 \includegraphics[width=3.5in, height=2.2in]{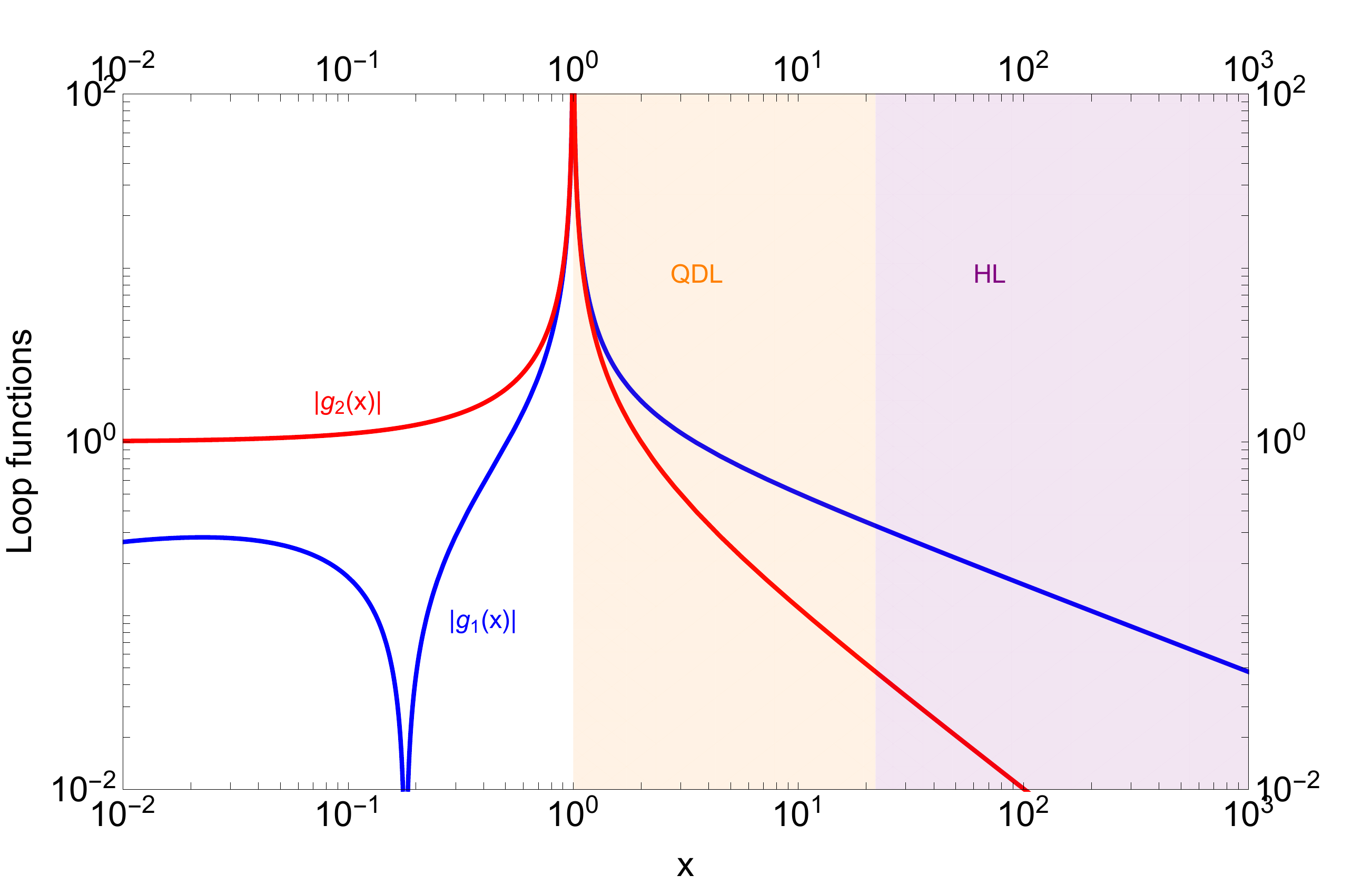}
 \caption{Upper panel: (Colour codes for the $N_{B-L}$ asymmetries are same as Fig.\ref{f1}): representative plots showing the validity of $N_1$ dominated scenario when the decay parameter of $N_2$ is weaker than the decay parameters of $N_1$ (left), and vice-versa (right). Bottom panel: Variation of the involved loop functions in the CP asymmetry parameters. The light violet region is the region where hierarchical scenario is valid in the model under consideration. 
 }\label{4plots}
\end{figure}

We first concentrate on the choice of RH neutrino mass hierarchy in this model. To find the minimum value of $M_2/M_1$, we generalise the procedure described in Sec.\ref{s3} and find that one may choose the RH neutrino mass hierarchy as mild as $M_2/M_1\sim 4.7$ for a perfectly valid $N_1{\rm DS}$\footnote{We have checked this using Eq.\ref{effi1} as well as Eq.\ref{fit}.}. In the upper panel of Fig.\,\ref{4plots}, we show the evolution of the $B-L$ asymmetry produced by both the RH neutrinos, in the two extreme cases of  $K_{1\tau}$ and  $K_{2\tau}$ (see right panel of Fig.\,\ref{fig6}). Note that, though for the first set of the decay parameters ($K_{2\tau}=5$ and $K_{1\tau}=25$), hierarchical $N_1$DS can be reproduced with $\delta_{12}\sim 1$, the second set ($K_{2\tau}=25$ and $K_{1\tau}=5$) requires a larger value of $\delta_{12}\sim 3.7$. For the first case, the $N_2$-washout is not strong enough to affect the asymmetry production by $N_1$ up to very low values of $\delta_{12}(\sim 1)$. Thus for $\delta_{12}\geq 1$, the final dynamics is governed only by the $N_1$-interactions. On the other hand, for the second case, the $N_2$-washout is much  stronger and it starts to reduce the magnitude of the asymmetry produced by $N_1$, unless one goes beyond $\delta_{12}\geq 3.7$. Henceforth, we designate $\delta_{12}=3.7$ as the critical point which separates the hierarchical (HL) and quasi-degenerate limit (QDL) of leptogenesis in $\rm CP^{\mu\tau}$ model. We use this mild hierarchy criteria, i.e., $M_2/M_1=4.7$ in the computation of leptogenesis for rest of the paper.

 Once we go from strong to a mild hierarchy, we immediately see an enhancement in the loop functions (cf. the bottom panel of Fig.\,\ref{4plots}). In hierarchical limit, it is sufficient to consider the enhancement in the function $g_1(x_{12}=M_2^2/M_1^2)$ which dominates  in $\varepsilon_{1\alpha}$.
 Due to this enhancement, the previously quoted lower bound on $M_1$ ($\sim 6\times 10^{10}$ GeV) \cite{cp3,cp16} gets lowered to $M_1^{\rm min}\sim 7.5\times 10^{9}$ GeV. Note that this can be further relaxed with the inclusion of flavour couplings, which tend to increase the efficiency of the asymmetry production. In addition, due to this choice of mild hierarchy $M_2$ would likely to be in the 2FR (the green rectangles in Fig.\,\ref{fig7}). However we stress that if one chooses a strong hierarchy, say $M_2/M_1=10^3$ (\cite{cp3,cp16,cplepto1}), $M_2$ is necessarily in the 1FR, if we take $M_1$ to be in the 2FR. Thus contribution from $M_2$ can be neglected since the total CP asymmetry vanishes in the unflavoured (1FR) regime. Therefore, the results obtained in the above references (for a pure $N_1$ domination) hold true.

\begin{figure}[!t]
 \includegraphics[scale=0.45]{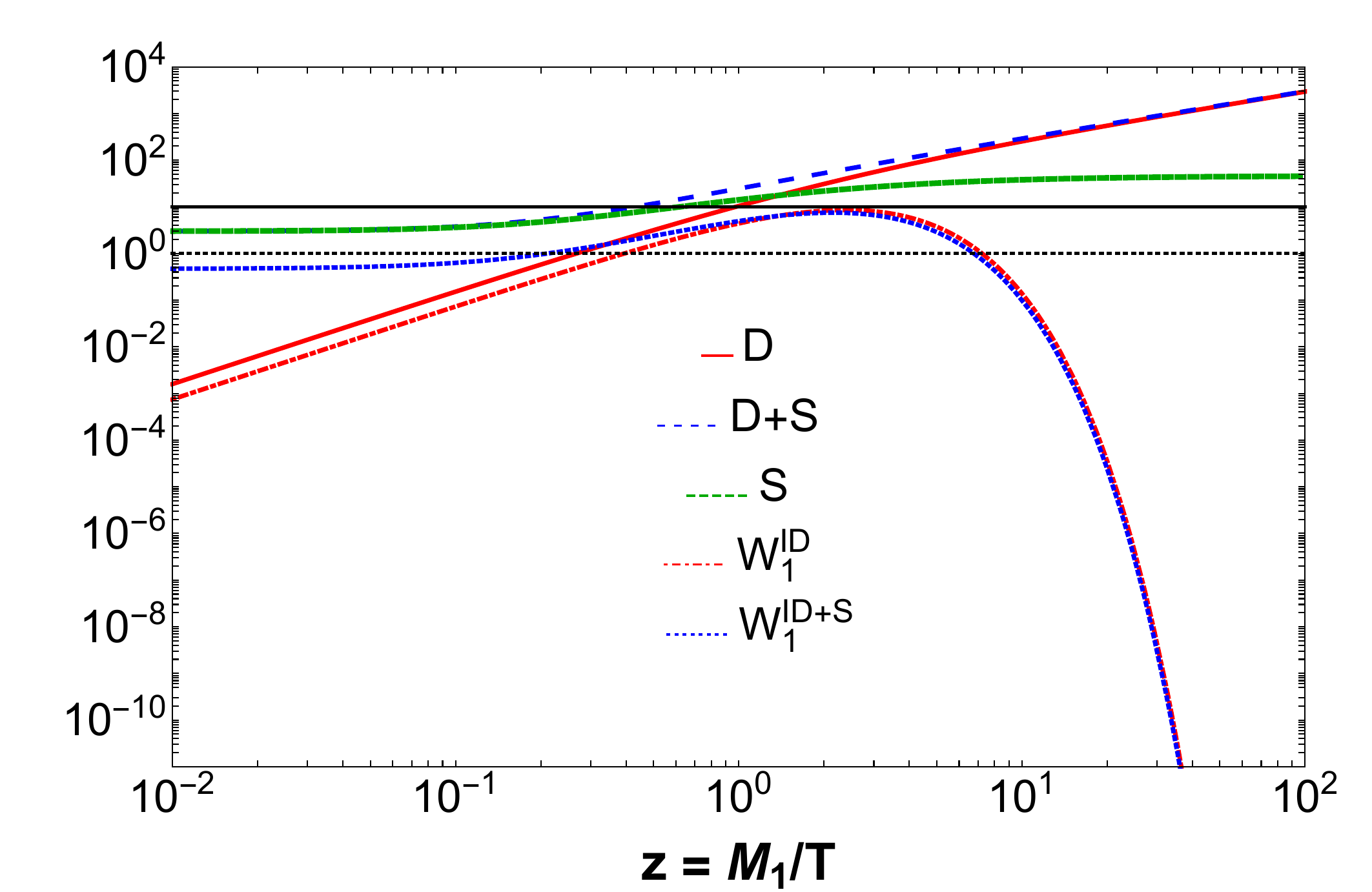}
 \caption {Comparison among rates of various processes involved in the leptogenesis. The horizontal black line is the rate of the $\tau$ charged lepton flavour interaction $\Gamma_\tau/Hz$ at the $N_1$ leptogenesis temperature $T\sim M_1\sim 10^{11}$ GeV. Domination of $\Gamma_\tau/Hz$ (over all the rates) has been considered to ensure a strongly decoherent picture for simplicity. }\label{4splots}
\end{figure}

 It is also worth mentioning that in this work, we consider a fully flavoured scenario where the charged lepton flavour interaction rate is dominant throughout the thermal history of the asymmetry production. Mathematically, this implies that the washout term $W(z^{max},K_1)<\Gamma_\tau/2Hz$, where $\Gamma_\tau$ is the $\tau$ interaction rate. This condition translates into
\bea
F_\tau\equiv \Gamma_\tau/2Hz_i=\frac{5\times ~10^{11}{\rm GeV}}{M_i}>W(z^{max}_i)\,,\label{coher}
\eea
where the washout term $W(z)$ contains inverse decays as well as dominant scattering rates. Notice that for a weak washout scenario, Eq.\,\ref{coher} is trivially satisfied. In that case, the washout terms never reach equilibrium and thus, for any value of $M_i<5\times 10^{11}$ GeV, the interaction rate $\Gamma_\tau$ is fast enough to break the coherence of the states produced by $N_i$. But for a strong washout, this is not the case since the washout term $W(z)\gg 1$. Thus, the masses for the RH neutrinos should be chosen carefully so that throughout the thermal history,  $\Gamma_\tau$ dominates over the relevant washout rates. Otherwise, one needs to take into account the off-diagonal terms of the density matrix\footnote{Note that thus the CP asymmetry parameter defined in Eq.\ref{pflp} appears in the `$\alpha\alpha$' (diagonal) term of the density matrix evolution equation \cite{Blanchet:2011xq,Dev:2014laa}. } that account for the coherence among the basis states \cite{Blanchet:2011xq,Moffat:2018wke,Dev:2014laa}. In this context, it is also worthwhile to recall Ref.\cite{q1,q2,q3,q4} that discuss leptogenesis using full quantum kinetic equations.

In the washout term, in addition to the inverse decay we include dominant $\Delta L=1$ scattering processes involving top quark. These processes include a combined contribution of the Higgs mediated $s$-channel ($N_i \ell\leftrightarrow qt$) and $t$-channel processes ($N_i q\leftrightarrow \ell t$). The relevant scattering rates for both the channels can be written as
\bea
S_{\phi i}^a=\frac{\Gamma_{\phi i}^a}{Hz}\,, \hspace{1.4cm}a=s,t\,.
\eea
The quantity $\Gamma_{\phi i}^a$ is related to the reaction density $\gamma_{\phi i}^a$ as $\Gamma_{\phi i}^a=\frac{\gamma_{\phi i}^a}{n_{N_i}^{\rm eq}}$, where for the reaction density of a generic $2\leftrightarrow 2$ process, one has the expression \cite{lep3}
\bea
\gamma(2\leftrightarrow 2)= \frac{g_x g_yT}{32\pi^4}\int ds s^{3/2}K_1(\sqrt{s}/T)\lambda \left(1,\frac{m_x^2}{s},\frac{m_y^2}{s}\right)\sigma(s)^a\,,
\eea
where $g_x$ and $g_y$ are initial state degrees of freedom, $s$ is the center of mass energy and the quantity $\lambda$ is given by
\bea
\lambda \left(1,\frac{m_x^2}{s},\frac{m_y^2}{s}\right)=\left(1-\frac{m_x^2}{s}-\frac{m_y^2}{s}\right)^2-4\frac{m_x^2m_y^2}{s^2}\,.
\eea
The washout for the $\Delta L=1$ term could be written as 
\bea
W_i^{\Delta L=1}=W_i^s+2W_i^t\,,
\eea
which are related to the scattering rate as 
\bea
W_i^s=\frac{N_{N_i}}{N_\ell^{\rm eq}} S_{\phi i}^s, W_i^t=\frac{N_{N_i}^{\rm eq}}{N_\ell^{\rm eq}} S_{\phi i}^t\,.
\eea
We compare the total washout term $W=W_i^{\rm ID}+W_i^{\Delta L=1}$ with the charged lepton interaction rate $F_\tau$. Given the ranges of the decay parameters, we find that  $M_1\sim 4\times 10^{10}$ GeV could be a safe value to circumvent the dominance of the washout processes over the charged lepton interaction\footnote{In the numerical computaion we use $M_\phi/M_1=10^{-5}$\cite{Luty:1992un,plum1}, where $M_\phi$ is the Higgs thermal mass needed to cut off the infrared divergences of $t$ channel process. For the the scattering cross sections please see Ref.\cite{plum2}. }. Notice that, for the mass window $M^{\rm max}\sim4\times 10^{10}~{\rm GeV}~\gtrsim M_1\gtrsim M_1^{\rm min}\sim  7.5\times 10^{9}$, the formulae we use in this paper are technically valid. While the inclusion of flavour couplings could lower the value of $M_1^{\rm min}$ (as already pointed out before), one may still go beyond $M^{\rm max}$ and opt for a diagonal density-matrix formalism. However, in that case one has to neglect the higher values of the decay parameters (i.e., there would be upper bound on the decay parameters) which are responsible for the dominance of washout terms over the charged lepton interactions. 
\begin{figure}[!t]
\includegraphics[scale=0.4]{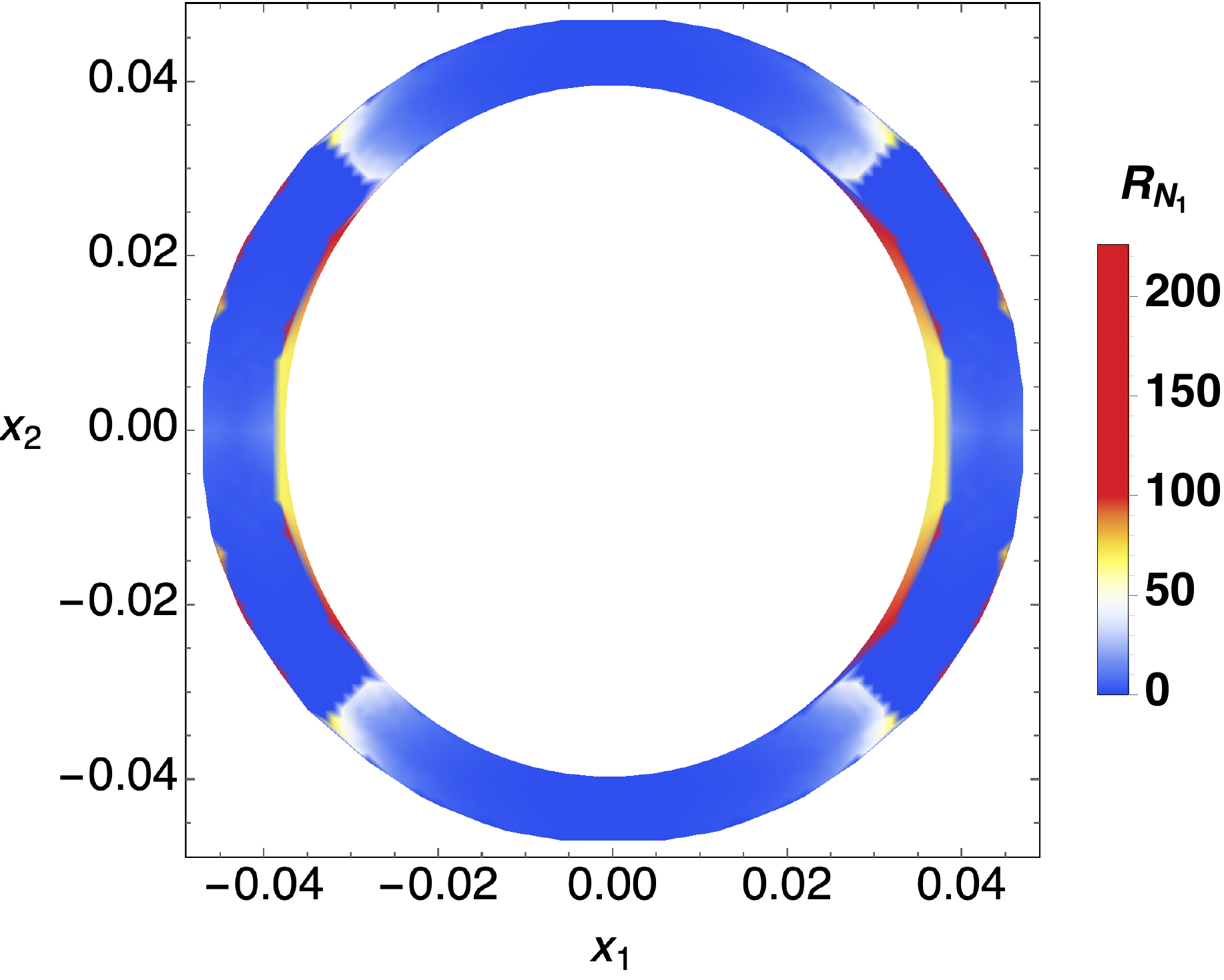} \includegraphics[scale=0.4]{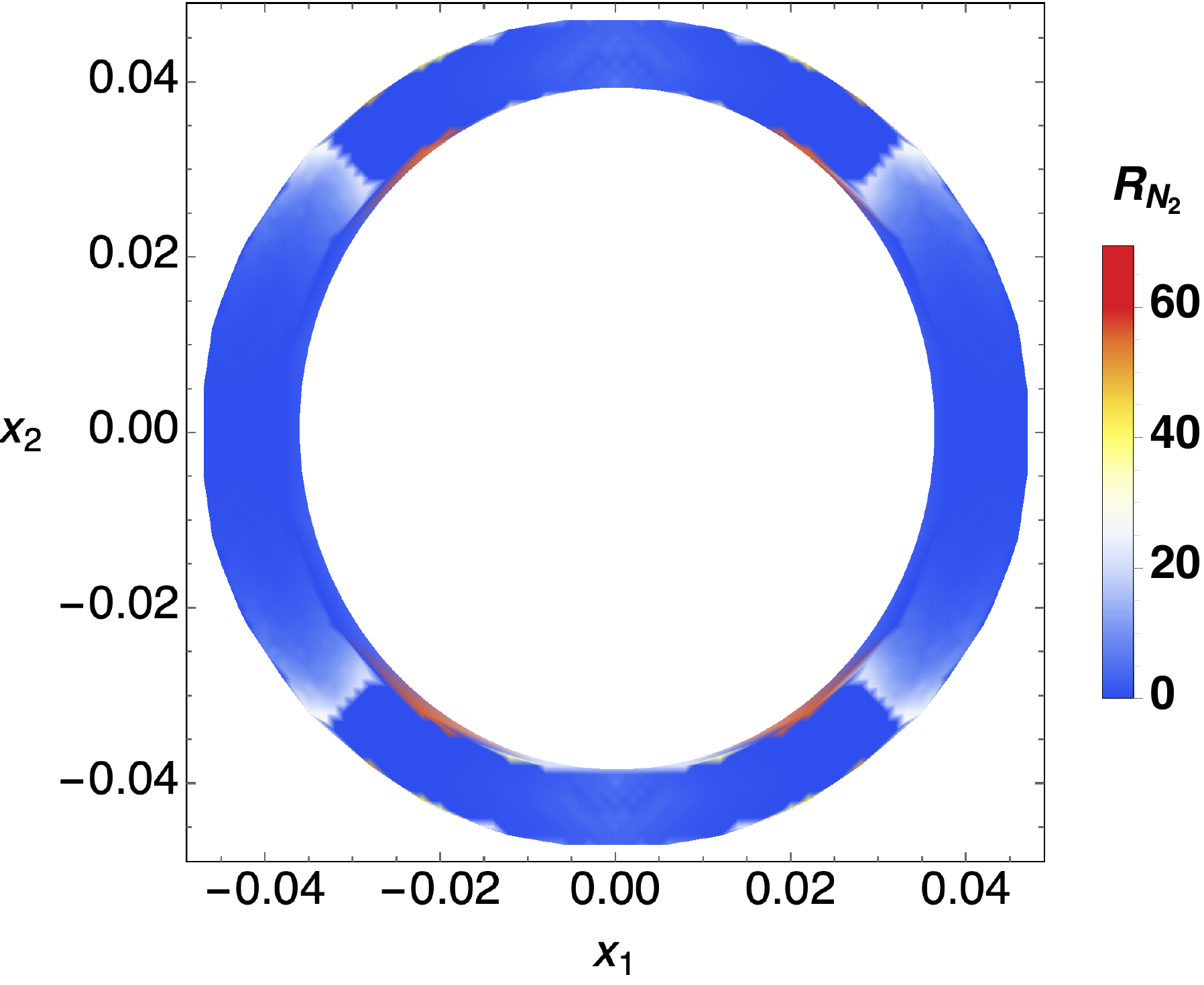}
\caption{Top panel: $R_{N_1}\equiv\eta_B^{N_1}/\eta_B^{N_2}$ with the the model parameters $x_1$ and $x_2$. Bottom panel: $R_{N_2}\equiv\eta_B^{N_2}/\eta_B^{N_1}$ with the the model parameters $x_1$ and $x_2$. All the plots are generated for $M_1=4\times 10^{10}$ GeV. Note that $x_{1,2}$ are dimensional quantities, and are plotted in units of $ \sqrt{\rm eV}$.}\label{domin}
\end{figure}
 In order to understand the contribution from $N_1$ and $N_2$ to $N_{B-L}$, we can write Eq.\,\ref{flaI} as
\bea
N_{B-L}=-\varepsilon_{1\tau}(\kappa_{1\tau}^\infty-\kappa_{1\tau^\perp}^\infty)-(1-p_{12}^\perp)\varepsilon_{2\mu}\kappa_{2\mu}^\infty=N_{B-L}^{N_1}+N_{B-L}^{N_2}
\eea
where $N_{B-L}^{N_1}$ is the contribution from $N_1$ and $N_{B-L}^{N_2}$ is the contribution from $N_2$. The ratios
\bea
R_{N_1}=\left | \frac{N_{B-L}^{N_1}}{N_{B-L}^{N_2}}\right |, \hspace{1cm}R_{N_2}=\left | \frac{N_{B-L}^{N_2}}{N_{B-L}^{N_1}}\right |
\eea
can be used to realize a particular $N_i$ domination quantitatively. 
We use the criteria $R_{N_i}>10$ to signify a particular $N_i$ domination. Notice from Fig.\,\ref{domin} that indeed both the $R$ parameters can have values $\gg 10$; also, in general, $R_{N_{1}} > R_{N_2}$. Thus lepton asymmetry produced by both the neutrinos can dominate for certain region of the parameter space. Quantitatively, 37$\%$ of the parameter space  favours a $N_1$ dominated scenario ($R_{N_1}>10$) and $26\%$ of the parameter space favours a $N_2$ dominated scenario ($R_{N_2}>10$). These percentages have been calculated by taking the ratios of the number of data points corresponding to $R_{N_i}>10$ and the total number of data points compatible with 3$\sigma$ neutrino oscillation data. We stress that the above quantification is valid for any arbitrary values of $M_1$ in the mass window $4\times 10^{10}~{\rm GeV}\gtrsim M_1\gtrsim M_1^{\rm min}$. 

\begin{figure}[!t]
\begin{center}
 \includegraphics[scale=0.15]{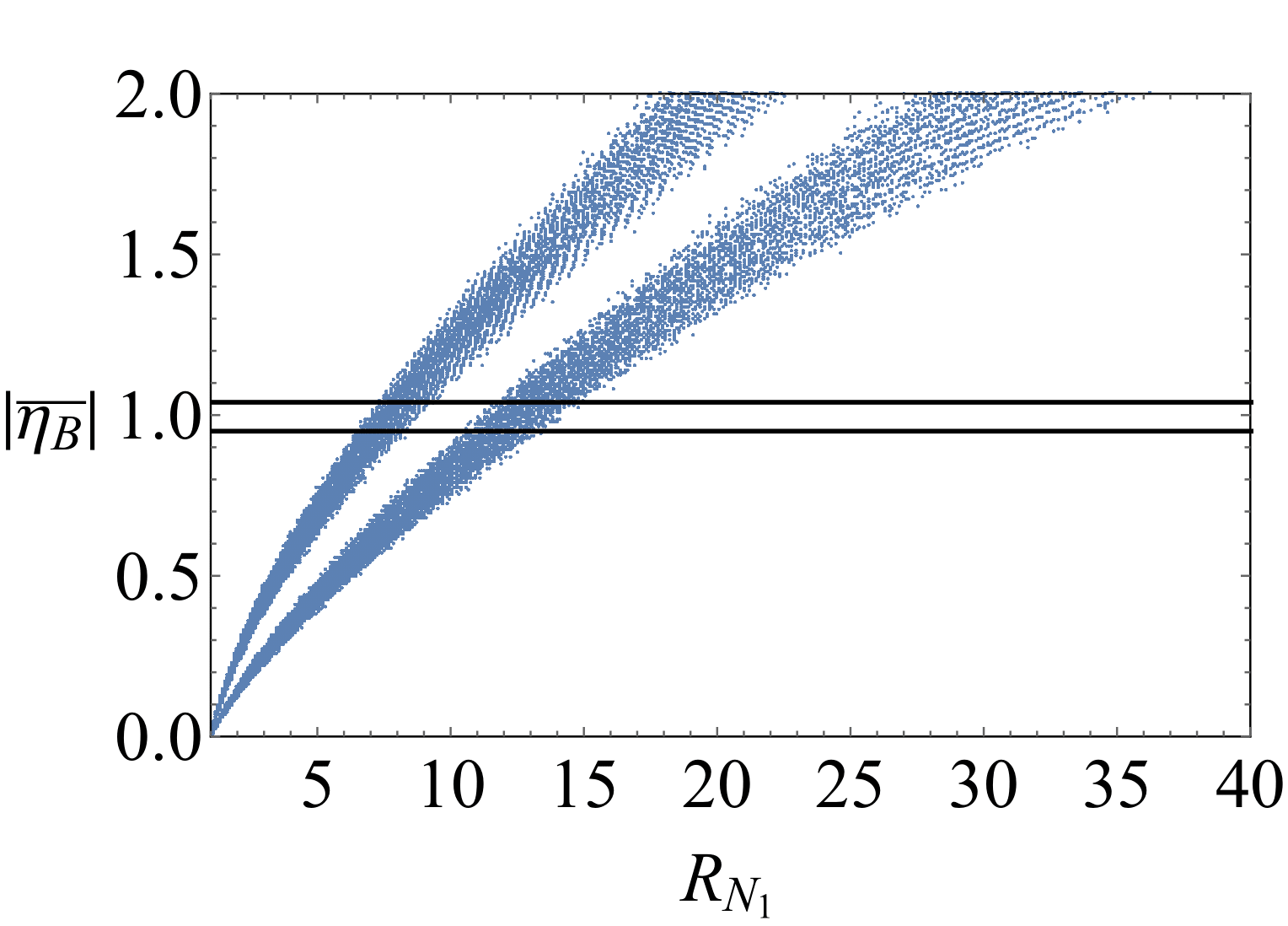} \includegraphics[scale=0.15]{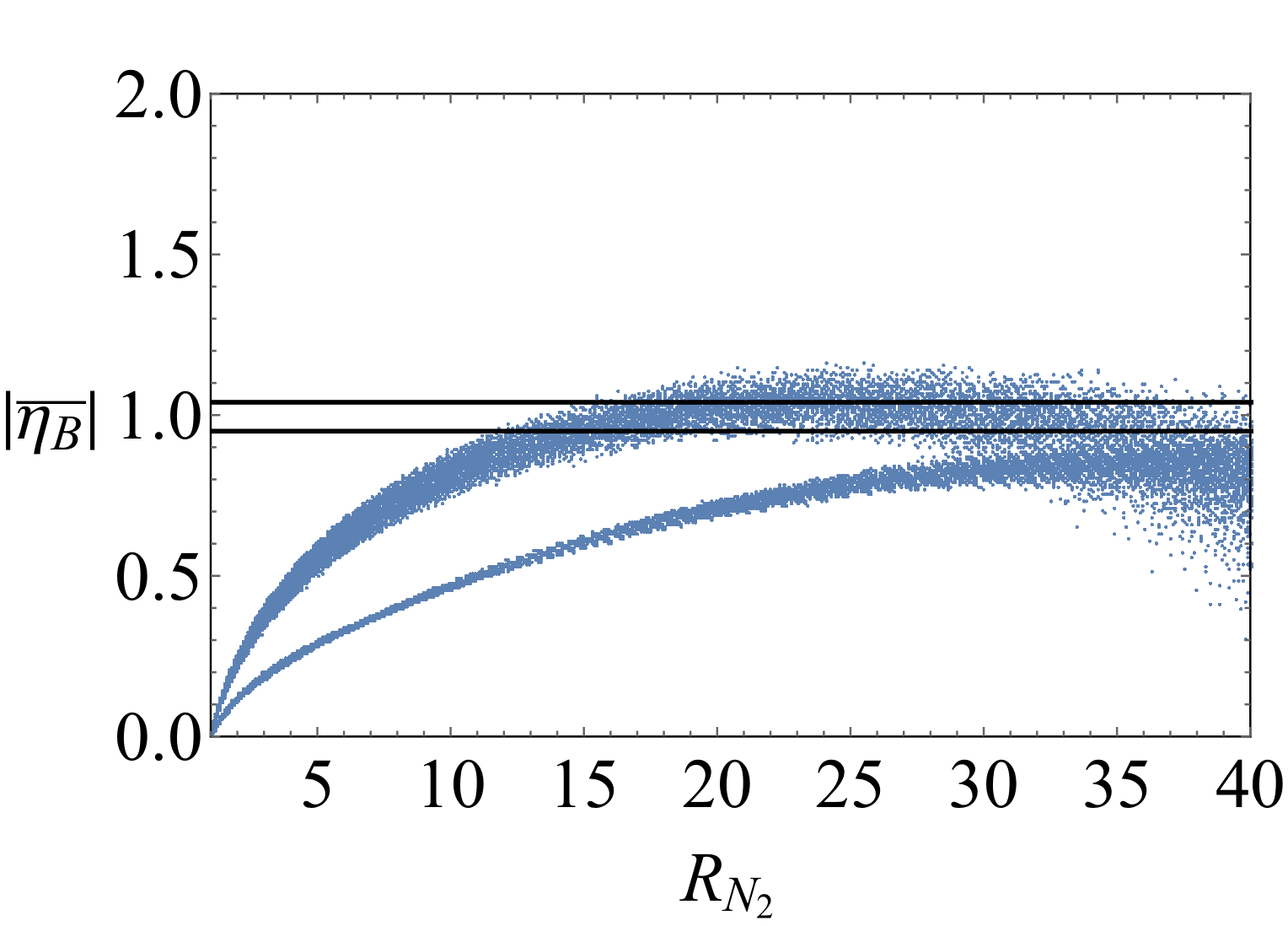}
 \caption{Normalised baryon to photon ratio with the $R$ parameters which quantify a particular $N_i$ dominance.}\label{ybrn}
\end{center}
\end{figure}

However,  the real challenge is now to check whether the parameters corresponding to $R_{N_1}>10$ or $R_{N_2}>10$ are able to reproduce the {\it observed range} of the baryon to photon ratio.  We checked that though the $N_1$ domination can be realized within the allowed mass window, $N_2$ domination can be realized marginally even if we take the maximal allowed value of $M_1^{\rm max}\sim 4\times 10^{10}~{\rm GeV}$.
In Fig.\,\ref{ybrn}, we plot the baryon to photon ratio normalised to $6.3\times 10^{-10}$ with the $R$ parameters for $M_1=7\times 10^{10}$ GeV.  
Note that, though this value of $M_1$ is beyond $M_1^{\rm max}$,  we do not lose any information on the RH neutrino masses by discarding the higher values of the decay parameters. Since higher values of $\eta_B$ correspond to lower values of the decay parameter, exclusion of higher values of the decay parameter implies truncating the lower portion of the parameter space (right hand side of Fig.\,\ref{ybrn}) in the $R_{N_2}-\overline{|\eta_{B}|}$ plane which is anyway much below $\overline{|\eta_{B}|}=1$. 

%%%%%%%%%%%%
%%%%%%%%%%%%
\subsection{$10^9\,{\rm GeV}<M_{2}<10^{12}\,{\rm GeV}$ and $M_{1}<10^{9}\,{\rm GeV}$}
In this section we give a qualitative picture of what happens in the case $10^9\,{\rm GeV}<M_{2}<10^{12}\,{\rm GeV}$ and $M_{1}<10^{9}\,{\rm GeV}$. Two important points should be stressed a priori. Firstly, since the mass of $N_1$ is much less than $10^9$ GeV, the CP asymmetry parameter $\varepsilon_{1\alpha}$ is highly suppressed and does not suffice to reproduce the correct baryon asymmetry \cite{ibarra}. One might wonder whether $N_2$ could produce a viable CP asymmetry or not. However, if we are in a two RH neutrino scenario (i.e., the third heavy neutrino does not couple to Higgs and leptons), the CP asymmetry parameter $\varepsilon_{2\alpha}$ (cf Eq.\,\ref{epn2}) is also proportional to $M_1$ and hence, suppressed by the small values of $M_1$. Therefore, in order to produce the correct amount of CP violation, one must need the $N_3$  to couple with $N_2$. 

Once $N_3$ is included in the discussion, we have more combinations of the RH mass spectrum on top of what has been shown in Fig.\,\ref{fig7}. However in this paper, we only consider the case $M_3>10^{12}$ GeV so that the asymmetry generated by $M_3$ vanishes (due to ${\rm CP^{\mu\tau}}$) and we have contributions from $N_2$ with $10^9$ GeV $<M_{2}<10^{12}$ GeV. Note that this mass spectrum \footnote{This is a very interesting mass spectrum for which a particular RH neutrino lies in a particular flavour regime, i.e., $M_3$ is in 1FR, $M_2$ is in 2FR and $M_1$ is in 3FR. This mass spectrum is often realized in SO(10) models\cite{Akhmedov:2003dg,DiBari:2008mp,Fong:2014gea}} ($M_3 \gg M_2 \gg M_1$), implies a strong hierarchical scenario. Thus, unlike the case discussed in the earlier section, any of the components of the asymmetry generated by $N_2$ does not escape $N_1$-washout since, for this mass spectrum of the RH neutrinos, $M_1$ is in the 3FR and the directions of $N_1$-washout coincide with that of the charged leptons. 
\begin{figure}[!t]
\begin{center}
\includegraphics[scale=0.55]{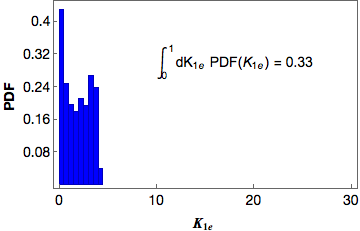} \includegraphics[scale=0.55]{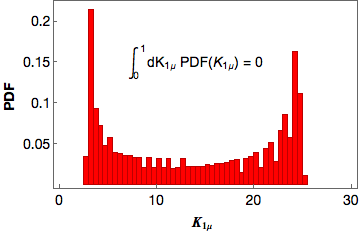}
\caption{Distributions of the flavoured decay parameters. The probability for the electron decay parameter $K_{1e}$ being less than 1 is almost $33\%$ which corresponds to the fact that asymmetry generated by $N_2$ is most likely to survive against $N_1$ washout in the electron flavour.}\label{dist}
\end{center}
\end{figure}

Following the above discussion, neglecting the contribution from $N_1$ and using Eq.\,\ref{fep5} -- Eq.\,\ref{fep7},  $B-L$ asymmetry parameter can now be written as 
\bea
N_{B-L}&=&-\varepsilon_{2\tau}\kappa_{2 \tau}^\infty e^{-3\pi K_{1\tau}/8}-\frac{K_{2\mu}}{K_{2\tau^\perp}}\varepsilon_{2\tau^\perp}\kappa_{2 \tau^\perp}^\infty e^{-3\pi K_{1\mu}/8}-\frac{K_{2e}}{K_{2\tau^\perp}}\varepsilon_{2\tau^\perp}\kappa_{2 \tau^\perp}^\infty e^{-3\pi K_{1e}/8}\,.\label{3flvasym}
\eea
 Note that each term in the RHS of Eq.\,\ref{3flvasym} contains the exponential washout factor involving the flavoured decay parameters. Thus strength of the $N_1$-decay parameters would finally decide whether the asymmetry generated by $N_2$ would survive against $N_1$-washout. Typically, $K_{1\alpha}<1$ is the condition for the washout processes to be considered ineffective (see \cite{DiBari:2005st,DiBari:2018fvo}), and thus  $P(K_{1\alpha}<1)$ is the probability for the asymmetry generated by $N_2$ to survive in the direction of $`\alpha$'. Given a general seesaw formula (constituent mass matrices are not subjected to any symmetry),  it has been shown for hierarchical light neutrinos that $P(K_{1e}<1)$ : $P(K_{1\mu}<1)$ : $P(K_{1\tau}<1) \simeq$ 0.36 : 0.058 : 0.067 $\simeq$ 6.2 : 1 : 1.15 \cite{DiBari:2018fvo}. For the ${\rm CP^{\mu\tau}}$ symmetric case, it is natural to infer that these probabilities would decrease, since in this case due to the imposed symmetry, there are now lesser number of parameters in the light neutrino mass matrix. For example., we compute these probabilities assuming hierarchical light neutrinos \footnote{ In our case assuming $N_3$ has Yukawa couplings $(m_D)_{3\alpha}$ which are similar order of magnitude as that of $N_1$ or $N_2$ so that in the seesaw light neutrino mass matrix $(m_D)_{3\alpha}$ is suppressed by the mass of $M_3$.} and in Fig.\,\ref{dist}, we show the corresponding distributions. It is evident that, though for the electron flavour we have $P(K_{1e}<1)\sim 0.33$, for the other two flavours (having same distribution due the $\mu\tau$ symmetry), the parameter space for $P(K_{1\mu,\tau}<1)$ closes.
 %%%
 %%`
 %
\begin{figure}[!t]
 \includegraphics[scale=0.3]{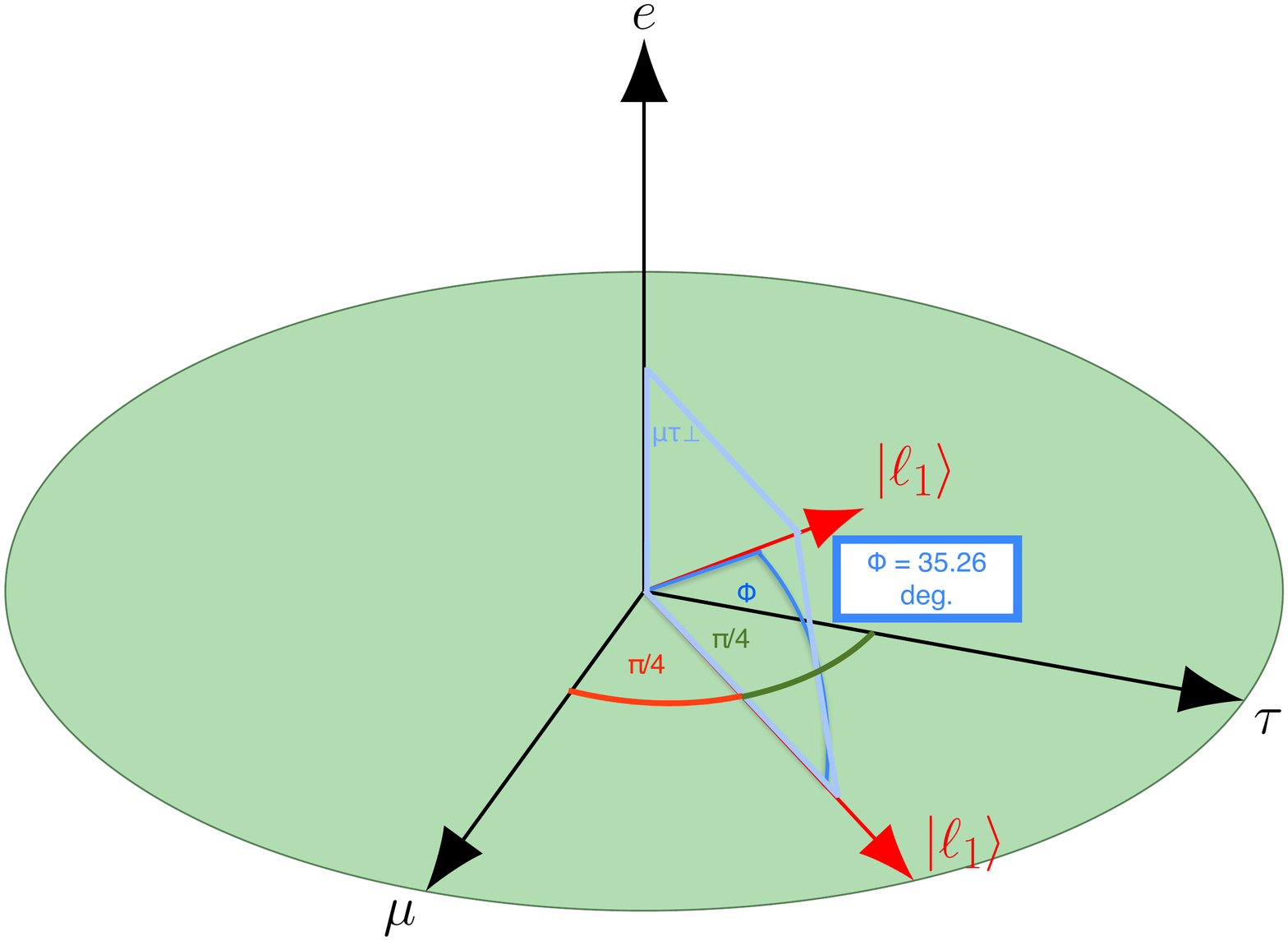} \includegraphics[scale=0.15]{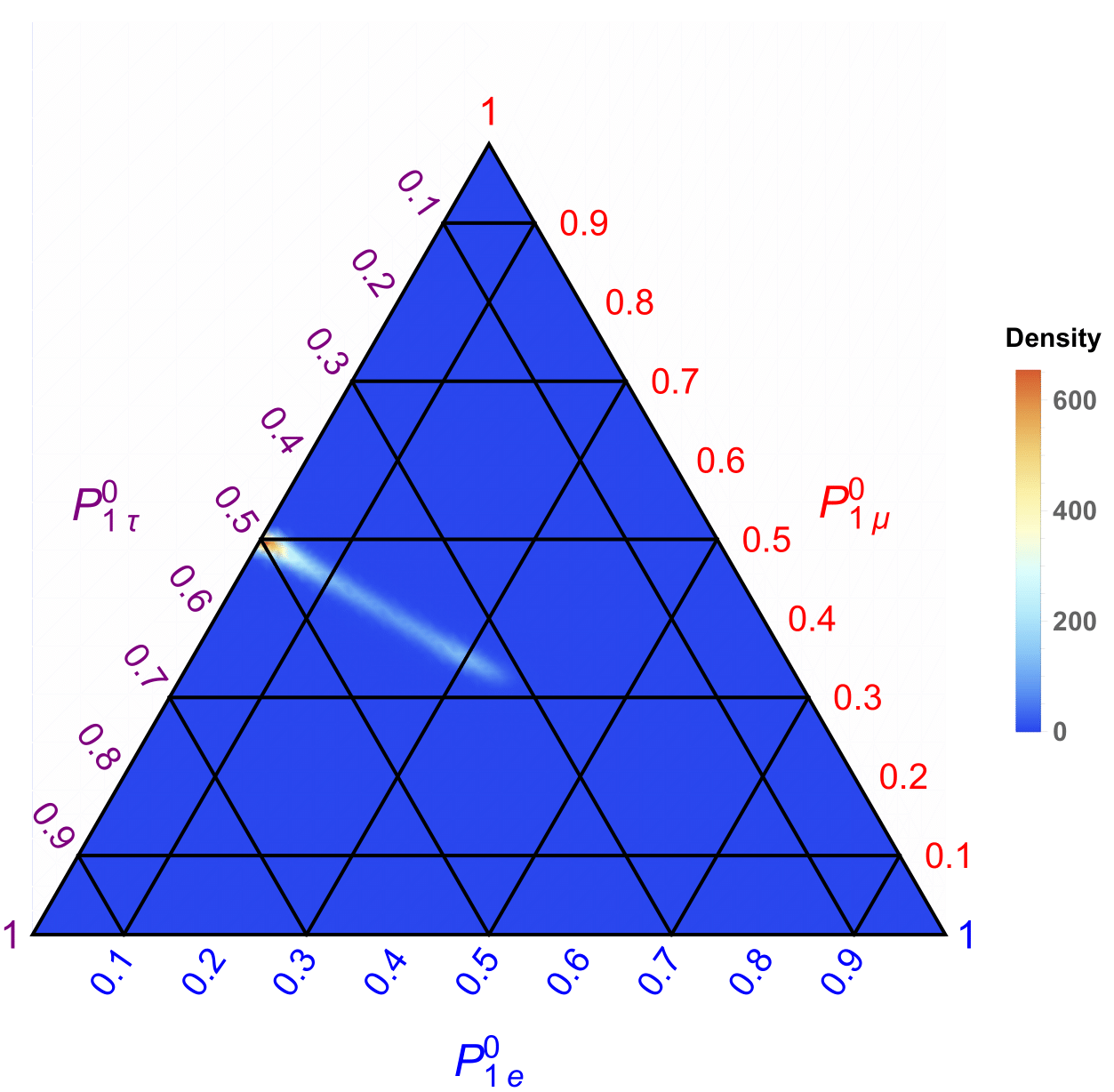}
\caption{Left: Possible range of the orientation of the state $\ket{\ell}_1$ in the ${\rm CP}$ symmetric model with hierarchical light neutrinos. Right: In the same model, ternary plot for the probabilities $P_{1\alpha}=K_{1\alpha}/\sum_{\alpha}K_{1\alpha}$ with corresponding densities.}\label{ori}
\end{figure}
%%%
%%%
This implies that  the asymmetry generated by $N_2$ would  survive in the electron flavour only. Note  that since  smaller values of $K_{1e}$ are most probable, there are more number of points for the smaller values of $P_{1e}=K_{1e}/K_1$. This implies the states $\ket{\ell}_1$  tend to lie on the $\mu-\tau$ plane.  The feature of getting mostly smaller values of $P_{1e}$ is quite generic\cite{DiBari:2018fvo}, but there is a clear difference between the general case and  $\rm CP^{\mu\tau}$. For the latter, the entire $\mu-\tau$ plane is not accessible to $\ket{\ell}_1$,  since  in this case  $P_{1\mu}=P_{1\tau}$,   and therefore the state $\ket{\ell}_1$ will have a definite direction ($45^0$ w.r.t $\mu$ or $\tau$ axis) on the $\mu-\tau$ plane.  In addition,  all possible  orientations of the $\ket{\ell}_1$ will lie on the plane $\mu\tau\perp$ as shown in the left panel of Fig.\ref{ori}. In the right panel, we show the triangle plot for the $P_{1\alpha}$ with corresponding densities. It is evident that the maximum dense region corresponds to $P_{1\mu}=P_{1\tau}=0.5$ and the probability densities can have values upto the the center of mass of the probability triangle, i.e., $P_{1e}$ : $P_{1\mu}$ : $P_{1\tau} \simeq $ 1 : 1 : 1. This  suggests that there would be an upper limit (in this model $\sim 35.26^0$) on the angle $\Phi$ which measures the angular deviation of the state $\ket{\ell}_1$ from the $\mu\tau$ plane as shown in Fig.\ref{ori}. 
%%%%%%%%%%%%%%%%%
%%%%%%%%%%%%%%%%%
\section{Summary}\label{s6}
%%%%%%%%%%%%%%%%%
%%%%%%%%%%%%%%%%%
In this work, we have performed a detailed study of the flavoured leptogenesis scenario in $\rm CP^{\mu\tau}$ symmetric neutrino mass models. We have shown how a mildly hierarchical leptogenesis ($M_2\simeq 4.7 M_1$) can be realized within the two flavour regime.  Within this class of models, even within the $N_1$-dominated scenario, the previously existing lower bound on $M_1$ can further be lowered approximately by an order of magnitude.
 Contrary to the previous works we have shown how in the two flavour regime, one can have a comparable parameter space for $N_2$- leptogenesis in addition to the standard $N_1$- leptogenesis. We have quantified the relevant mass scales of the RH neutrinos for a $N_i$-leptogenesis to dominate.

Taking the appropriate flavour effects into account, we have argued that the standard hierarchical $N_1$-dominated scenario is valid only for the mass window ($M_1^{\rm max}$)$\sim 4\times10^{10}$ GeV $>M_1>(M_1^{\rm min})\sim 7.5\times10^{9}$ GeV. 
Else, if the mass of $N_1$ goes beyond $M_1^{\rm max}$, there is a substantial amount of parameter space for which a $N_2$-dominated scenario could also be realized.
We have considered other mass spectra of the heavy neutrinos for which the lepton asymmetry generated by $N_2$ in two flavour regime faces washout by $N_1$ in the three flavour regime. For a hierarchical light neutrino mass spectrum, we have demonstrated that approximately one third of the parameter space allows an {\it electron-flavoured} $N_2$-leptogenesis to be realized.

The possibility of having a mildly hierarchical leptogenesis opens up several interesting avenues. With this detailed work, we hope to elucidate some aspects of this involved problem. Certainly, inclusion of several other effects, e.g., consideration of flavour couplings, quantum corrections to the neutrino  parameters would improve  the results presented in this paper. We plan to include these effects in a future work.

%%%%%%%%%%
%%%%%%%%%%
\section*{Acknowledgement}
We would like to thank Pasquale Di Bari for many useful discussions regarding leptogenesis. RS is supported by a Newton International Fellowship (NF 171202) from Royal Society (UK) and SERB (India). MS acknowledges support from the National Science Foundation, Grant PHY-1630782, and to the Heising-Simons Foundation, Grant 2017-228.

{}
\end{document}